\documentclass[journal]{IEEEtran}
\usepackage{cite}
\usepackage{amsmath,amssymb,amsfonts}
\usepackage{algorithmic}
\usepackage{graphicx}
\usepackage{textcomp}
\usepackage{times}
\usepackage{cite}
\usepackage{amssymb}
\usepackage{amsmath,epsfig}
\usepackage{subfigure}
\usepackage{float}
\usepackage{siunitx}
\usepackage{array}
\usepackage{setspace}
\usepackage{indentfirst}
\usepackage{algorithm}
\usepackage{algorithmic}
\usepackage{bm}
\usepackage{caption}
\usepackage{graphicx,xcolor}
\usepackage{color}
\usepackage{stfloats}

\renewcommand{\vec}[1]{\mathbf{#1}}

\DeclareMathAlphabet{\mathpzc}{OT1}{pzc}{m}{it}
\makeatletter
\newcommand*{\rom}[1]{\expandafter\@slowromancap\romannumeral #1@}
\makeatother
\usepackage{float}
\usepackage{multicol}

\usepackage{hyperref}
\usepackage{adjustbox}
\usepackage{multirow}

\begin{document}
\thispagestyle{empty}
\onecolumn
\begin{Large}IEEE Copyright Notice:\end{Large}

\vspace{1cm}
\textcopyright 2020 IEEE. Personal use of this material is permitted. Permission from IEEE must be obtained for all other uses, in any current or future media, including reprinting/republishing this material for advertising or promotional purposes, creating new collective works, for resale or redistribution to servers or lists, or reuse of any copyrighted component of this work in other works.

\vspace{1cm}
This work has been submitted to the IEEE for possible publication. Copyright may be transferred without notice, after which this version may no longer be accessible.

\newpage

\twocolumn
\setcounter{page}{1}

\title{Learning stochastic object models from medical~imaging measurements using Progressively-Growing~AmbientGANs}

\author{Weimin Zhou,
        Sayantan Bhadra,
        Frank J. Brooks, 
        Hua Li,~%
        and Mark A. Anastasio
        \thanks{
This work was supported in part by NIH Awards EB020604, EB023045, NS102213, EB028652, CA233873 and CA223799, and NSF Award DMS1614305. (Corresponding author: Mark A. Anastasio)}
\thanks{Weimin Zhou and Sayantan Bhadra are, respectively, with the Department 
of Electrical and Systems Engineering and the Department 
of Computer Science and Engineering, Washington University in St. Louis, St. Louis,
MO, 63130 USA e-mail: \{wzhou24, sayantanbhadra\}@wustl.edu.}
\thanks{Frank J. Brooks, Hua Li, and Mark A. Anastasio are with the Department 
of Bioengineering, University of Illinois at Urbana-Champaign, Urbana,
IL, 61801 USA e-mail: \{fjb, huali19, maa\}@illinois.edu. Hua Li is also with the Carle Foundation Hospital, Urbana, IL 61801 USA}
\vspace{-0.6cm}
}
\maketitle

\begin{abstract}
It has been advocated that medical imaging systems and reconstruction algorithms should be assessed and optimized by use of objective measures of image quality that quantify the performance of an observer at specific diagnostic tasks. One important source of variability that can significantly limit observer performance is variation in the objects to-be-imaged. This source of variability can be described by stochastic object models (SOMs). A SOM is a generative model that can be employed to establish an ensemble of to-be-imaged objects with prescribed statistical properties. In order to accurately model variations in anatomical structures and object textures, it is desirable to establish SOMs from experimental imaging measurements acquired by use of a well-characterized imaging system. Deep generative neural networks, such as generative adversarial networks (GANs) hold great potential for this task. However, conventional GANs are typically trained by use of reconstructed images that are influenced by the effects of measurement noise and the reconstruction process. To circumvent this, an AmbientGAN has been proposed that augments a GAN with a measurement operator. However, the original AmbientGAN could not immediately benefit from modern training procedures, such as progressive growing, which limited its ability to be applied to realistically sized medical image data. To circumvent this, in this work, a new Progressive Growing AmbientGAN (ProAmGAN) strategy is developed for establishing SOMs from medical imaging measurements. Stylized numerical studies corresponding to common medical imaging modalities are conducted to demonstrate and validate the proposed method for establishing SOMs.
\end{abstract}


\section{Introduction}
\label{sec:introduction}
\IEEEPARstart{C}{omputer}-simulation remains an important approach for the design and optimization
of imaging systems. Such approaches can permit the exploration, refinement, and assessment
 of a variety of system designs that would be infeasible through experimental studies alone.
In the field of medical imaging, it has been advocated that imaging systems
 and reconstruction algorithms
 should be assessed and optimized
by use of objective measures of image quality (IQ) that quantify the performance of an observer at specific diagnostic tasks \cite{myers1993rayleigh,wagner1985unified,barrett1993model,barrett2013foundations,anastasio2010analysis}.
To accomplish this,  all sources of variability in the measured data should  be
accounted for.
One important source of variability that can significantly limit observer performance
is variation in the objects to-be-imaged \cite{rolland1992effect}.
This source of variability can be described by stochastic object models (SOMs)
 \cite{kupinski2003experimental}.
A SOM is a  generative model that can be employed to produce an ensemble of to-be-imaged objects 
that possess prescribed statistical properties.

Available SOMs include  texture models of mammographic images with
clustered lumpy backgrounds \cite{bochud1999statistical}, simple lumpy background models
\cite{rolland1992effect}, and more realistic 
anatomical phantoms that can be randomly perturbed \cite{segars2008realistic}.
A variety of other computational phantoms~\cite{segars2002study,xu2014exponential,zankl2010gsf,collins1998design,caon2004voxel, zu2005vip, segars2008realistic,li2009methodology},
either voxelized or mathematical, have been proposed for medical imaging simulation,
 aiming to provide a practical solution to characterize object variability.
However, the majority of these were established by use of image data corresponding to a few subjects.
Therefore, they may not accurately describe the statistical properties
of the  ensemble of objects that is relevant to an imaging system optimization task. 
 A variety of anatomical shape models have also been proposed to describe both the common geometric features
and the geometric variability among instances of the population for shape
 analysis applications~\cite{cootes1995active,cootes2015active,heimann2009statistical,ferrari2010images,shen2012detecting,gordillo2013state,
tomoshige2014conditional,ambellan2019automated}.
To date, these have not been systematically explored for the purpose of
 constructing SOMs that capture realistic anatomical variations
for use in imaging system optimization.

In order to establish SOMs that capture realistic textures and anatomical variations,
it is desirable to utilize experimental imaging data.
By definition, however, SOMs should be independent of the imaging system, measurement noise
 and any reconstruction method employed. In other words, they should provide an \emph{in silico}
representation of the ensemble
of objects to-be-imaged and not estimates of them that would be indirectly measured
or computed by imaging systems.  To address this need, 
Kupinski \emph{et al.} \cite{kupinski2003experimental}
proposed an explicit generative model for describing object statistics
 that was trained by use of noisy imaging measurements and a computational model of
a well-characterized imaging system~\cite{kupinski2003experimental}.
However,
applications of this method have been limited to situations
 where the characteristic function of the corresponding imaging measurements can be analytically determined,
such as with lumpy and clustered lumpy object models\cite{kupinski2005small,bochud1999statistical}.
As such, there remains a need
to generalize the method so that anatomically realistic and more complicated SOMs
can be established from experimental imaging measurements.

Deep generative neural networks, such as generative adversarial networks
 (GANs) \cite{goodfellow2014generative}, hold great potential for establishing SOMs
that describe discretized objects.
However, conventional GANs are typically trained by use of reconstructed
 images that are influenced by the effects of measurement noise
 and the reconstruction process.
 To circumvent this, an AmbientGAN has been proposed \cite{bora2018ambientgan} that augments
 a GAN with a measurement operator. This permits
 a generative model that describes object randomness 
 to be learned from indirect and noisy measurements of the objects
themselves.
In a preliminary study, the AmbientGAN was explored
 for the establishing SOMs from imaging measurements for use in optimizing imaging systems \cite{zhou2019learning}.
 However, similar to conventional GANs, the process of training AmbientGANs
is inherently unstable.
Moreover, the original AmbientGAN cannot immediately benefit from
 robust GAN training procedures, such as progressive growing\cite{karras2017progressive},
 which limits its ability to synthesize high-dimensional images that depict
objects of interest in medical imaging studies.

In this work, a new AmbientGAN approach is proposed that permits the
utilization of the progressive growing strategy for training.
In this way, SOMs can be established from noisy imaging measurements that can yield
high-dimensional images that depict objects.
The new approach, referred to as a Progressive Growing AmbientGAN (ProAmGAN),  
can utilize the progressive growing training strategy due to 
augmentation of the conventional AmbientGAN architecture with an image
reconstruction operator.
Stylized numerical studies corresponding to X-ray computed tomography (CT) and
magnetic resonance imaging (MRI) are
conducted to investigate the proposed ProAmGAN for establishing SOMs.
Preliminary validation studies are presented that utilize standard quantitative
measures  for evaluating GANs and also objective measures based on signal detection
performance.

The remainder of this paper is organized as follows. 
In Sec.~\ref{sec:bkgd}, previous works on learning SOMs
 by employing characteristic functions and AmbientGANs 
are summarized.  
The progressive growing training strategy for GANs is also reviewed.
The proposed ProAmGAN for learning SOMs from noisy
 imaging measurements is described in Sec.~\ref{sec:ProAmGAN}.
Sections \ref{sec:num} and \ref{sec:result} describe the numerical studies and results
that demonstrate the ability of the ProAmGAN to learn SOMs from
stylized X-ray CT and MRI measurements.
Finally,  a discussion and summary of the work is presented in
Sec.~\ref{sec:conclusion}.

\section{Background}
\label{sec:bkgd}

Consider a discrete-to-discrete (D-D) description of 
a linear imaging system given by~\cite{barrett2013foundations}:
\begin{equation}\label{eq:imaging}
    \mathbf{g} = \mathbf{H}\mathbf{f} + \mathbf{n},
\end{equation}
\noindent where $\mathbf{g} \in \mathbb{R}^M$ is a vector that describes the measured image data,
$\mathbf{f} \in \mathbb{R}^N$ denotes the finite-dimensional representation of the object being imaged, 
$\mathbf{H}\in \mathbb{R}^{M\times N}$ denotes a D-D imaging operator $\mathbb{R}^N\rightarrow \mathbb{R}^M$ that maps an object in the Hilbert space $\mathbb{U}$ to the measured discrete data in the Hilbert space $\mathbb{V}$, 
and the random vector $\mathbf{n} \in \mathbb{R}^M$ denotes the measurement noise. 
Below, the imaging process described in Eq.~(\ref{eq:imaging}) is denoted as: $\mathbf{g} = \mathcal{H}_{\mathbf{n}}(\mathbf{f})$.  
It is assumed that the D-D imaging model is a sufficiently accurate representation
of the true continuous-to-discrete (C-D) imaging model that describes a digital imaging system and the impact of model error will be neglected.
When optimizing imaging system performance by use of objective measures of IQ, all sources
of randomness in $\mathbf{g}$ should be considered. In diagnostic imaging applications,
object variability is an important factor that limits observer performance.
In such applications, the object  $\mathbf{f}$ should be described as a random vector
that is characterized by a multivariate probability density function (PDF)
 $\textrm{pr}(\mathbf{f})$ that specifies the
statistical properties of the ensemble of objects to-be-imaged.

Direct estimation of $\textrm{pr}(\mathbf{f})$ is rarely tractable in medical imaging applications due 
to the high dimensionality of $\mathbf{f}$.
To circumvent this difficulty, a parameterized generative model, referred to throughout this work as
a SOM,  can be introduced and established
by use of an ensemble of experimental measurements. The generative model can be explicit
or implicit.
Explicit generative models seek to approximate $\textrm{pr}(\mathbf{f})$, or equivalently, its
characteristic function,  from which samples
$\mathbf{f}$ can subsequently be drawn. 
On the other hand, implicit generative models do not seek to estimate $\textrm{pr}(\mathbf{f})$ directly,
 but rather define a stochastic process that seeks to draw samples
from $\textrm{pr}(\mathbf{f})$ without having to explicitly specify it.  Variational autoencoders and GANs are examples
 of  explicit and implicit generative models, respectively, that have
been actively explored \cite{goodfellow2016deeplearning}.
Two previous works that sought to learn SOMs from noisy and indirect imaging measurements  by use of explicit and implicit generative
models are presented below.

\subsection{Establishing SOMs by use of explicit generative modeling: Propagation
of characteristic functionals}
\label{ssec:CFforSOM}

The first method to learn SOMs from imaging measurements was introduced by
Kupinski \textit{et al.}~\cite{kupinski2003experimental}.
In that work, a C-D imaging model was considered in which a function that describes the object is mapped to a finite-dimensional image vector
$\mathbf{g}$.
For C-D operators, it has been demonstrated that 
 the characteristic functional (CFl) describing the object 
 can be readily related to the characteristic function (CF) 
of the measured data vector $\mathbf{g}$ \cite{clarkson2002transformation}.
 This provides  a
relationship between the PDFs of the object and measured image data.
In their method, an object that was parameterized by
the vector $\mathbf{\Theta}$ was considered and
analytic expressions for the CFl were utilized.
Subsequently, by use of the known imaging operator and noise model,
  the corresponding CF  was computed. The vector $\mathbf{\Theta}$
was estimated by 
 minimizing the discrepancy between this model-based CF and
an empirical estimate of the CF computed from an ensemble of noisy
imaging measurements. From the estimated CFl,
an ensemble of objects could be generated.
This method was applied to establish SOMs where the CFl of the object can
be analytically determined. 
Such cases include the
lumpy object model\cite{kupinski2005small} and
 clustered lumpy object model \cite{bochud1999statistical}.
The applicability of the method to more complicated object
models  remains unexplored.

\subsection{Establishing SOMs by use of implicit generative
modeling: GANs and AmbientGANs}
\label{ssec:AmGANforSOM}

Generative adversarial networks (GANs)~\cite{goodfellow2014generative, arjovsky2017towards,arora2017do,denton2015deep,radford2015unsupervised,
salimans2016improved, li2019misGAN, shrivastava2015learning, arjovsky2017wasserstein, gulrajani2017improved, brock2018large} 
 are implicit generative models that have been actively explored to
learn the statistical properties of ensembles of images and generate new images
that are consistent with them.
A traditional GAN consists of two deep neural networks - a  generator and a discriminator. 
The generator is jointly trained with the discriminator through an adversarial process. 
During its training process, the generator is trained to map random low-dimensional latent vectors
to higher dimensional images that represent samples from the distribution of training images.
The discriminator is trained to distinguish the generated, or synthesized, images from
 the actual training images. These are often referred to as the ``fake" and ``real" images in the
GAN literature. 
Subsequent to training, the discriminator is discarded and the generator and associated
latent vector probability distribution form as an implicit
generative model that can sample from the data distribution to produce new images.
However, images produced by  imaging systems are contaminated by measurement
noise and potentially an image reconstruction process.  Therefore, GANs trained
directly on images do not generally represent SOMs because they do not
characterize object variability alone.

\begin{figure}[t]
\centering
\includegraphics[width=0.7\linewidth]{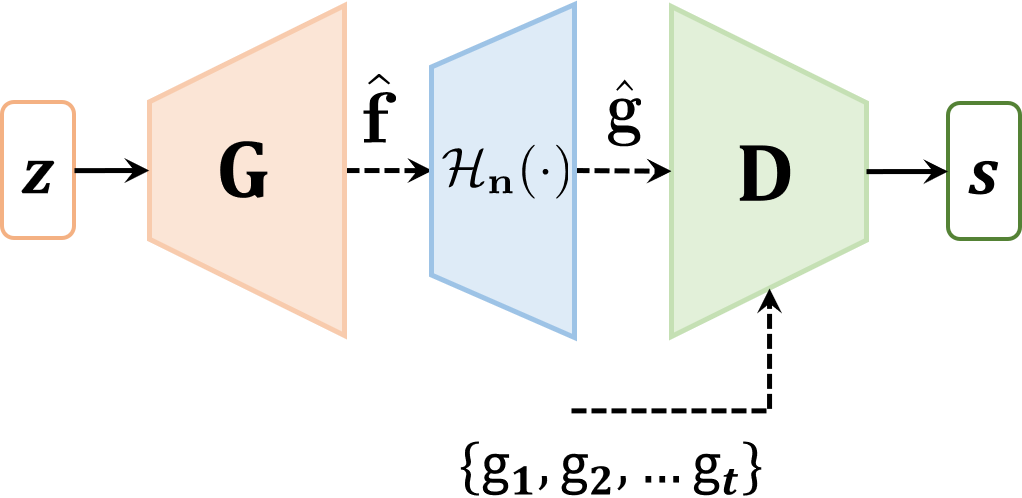}
\caption{An illustration of the AmbientGAN architecture. The generator $\mathbf{G}$ is trained to generate objects, which are subsequently employed to simulate measurement data. The discriminator $\mathbf{D}$ is trained to distinguish ``real" measurement data to the ``fake" measurement data that are simulated by use of the generated objects.}
\label{fig:arc_AGAN}
\vspace{-0.3cm}
\end{figure}

An augmented GAN architecture named AmbientGAN has been proposed~\cite{bora2018ambientgan}
that enables learning an SOM from noisy indirect measurements of an object.
As shown in Fig.~\ref{fig:arc_AGAN}, the AmbientGAN architecture includes
 the measurement operator  $\mathcal{H}_{\vec{n}}$, defined in Eqn.\ (\ref{eq:imaging}),
 into the traditional GAN framework.
During the AmbientGAN training process,
the generator is trained to map a random vector $\mathbf{z} \in \mathbb{R}^{k}$ described by a latent probability distribution to a generated object
$\hat{\mathbf{f}} = G(\mathbf{z}; \mathbf{\Theta}_{G})$, 
where $G: \mathbb{R}^{k} \rightarrow \mathbb{R}^N$ represents the generator network
 that is parameterized by a vector of trainable parameters $\mathbf{\Theta}_G$.
Subsequently, the corresponding simulated imaging measurements are computed as
 $\hat{\vec{g}} = \mathcal{H}_\vec{n}(\hat{\vec{f}})$.
The discriminator neural network $D: \mathbb{R}^{N} \rightarrow \mathbb{R}$,
which is parameterized  by the vector $\mathbf{\Theta}_D$,
is trained to distinguish the real and simulated imaging measurements by mapping them to real-valued scalar $s$.
The adversarial training process 
 can be represented by the following two-player minimax game\cite{goodfellow2014generative}:
\begin{equation} \label{eq:AGAN}
\begin{split}
\min_{\mathbf{\Theta}_G} \max_{\mathbf{\Theta}_D} V(D,G) = & {E_{\vec{g}\sim p_{\vec{g}}}} [l\left(D(\vec{g}; \mathbf{\Theta}_D)\right)]\\
 &+ {E_{\hat{\vec{g}} \sim p_{\hat{\vec{g}}}}} [l(1- D\left( \hat{\vec{g}}; \mathbf{\Theta}_D \right) )],
\end{split}
\end{equation}
where $l(\cdot)$ represents a loss function.  
When the distribution of objects $\mathrm{pr}(\vec{f})$ uniquely
 induces the distribution of imaging measurements $\mathrm{pr}(\vec{g})$, i.e., when the 
imaging operator is injective, and the minimax game achieves the global optimum, 
the trained generator can be employed to produce object samples drawn from $\mathrm{pr}(\vec{f})$~\cite{goodfellow2014generative, bora2018ambientgan}.   

Zhou \textit{et al.} have demonstrated the ability of the AmbientGAN to learn a
simple SOM corresponding to a lumpy object model that could be 
employed to produce small ($64\times 64$) object samples \cite{zhou2019learning}.
However,
adversarial training is known to be unstable
and the use of AmbientGANs to establish realistic and large-scale SOMs has, to-date, been limited.

\subsection{Progressively-Growing GAN Training Strategy}
\label{ssec:ProGAN}

A novel training strategy for GANs---progressive growing of GANs (ProGANs)---has been recently developed to improve the stability of the GAN training process~\cite{karras2017progressive} 
and hence the ability to learn  generators that sample
from distributions of high-resolution images.
GANs are conventionally trained directly on full size images through the entire training process.
 In contrast, ProGANs adopt a multi-resolution approach to training.
Initially, a generator and discriminator are trained by use of down-sampled (low resolution) training
images.
During each subsequent training stage, higher resolution versions of the original training images are employed
to train progressively deeper discriminators and generators, continuing until a final version of the
  generator
is trained by use of the original high-resolution images.
While this progressively growing training strategy has found widespread success with conventional GANs,
as described below,
it cannot generally be employed with AmbientGANs.
A solution to this problem is described next.

\section{Establishing SOMs by use of Progressively-Growing AmbientGANs}
\label{sec:ProAmGAN}

As discussed above, AmbientGANs enable the learning of SOMs from noisy imaging measurements but can be difficult to train, 
while ProGANs can be stably trained and established by use of higher-dimensional
image data that are generally affected by noise and the image formation process.
Below, a novel strategy, Progressively Growing AmbientGANs (ProAmGANs), is proposed 
to enable progressive growing of AmbientGANs for  learning
realistic SOMs from noisy and indirect imaging measurements. 

\begin{figure}[H]
\centering
 \includegraphics[width=\linewidth]{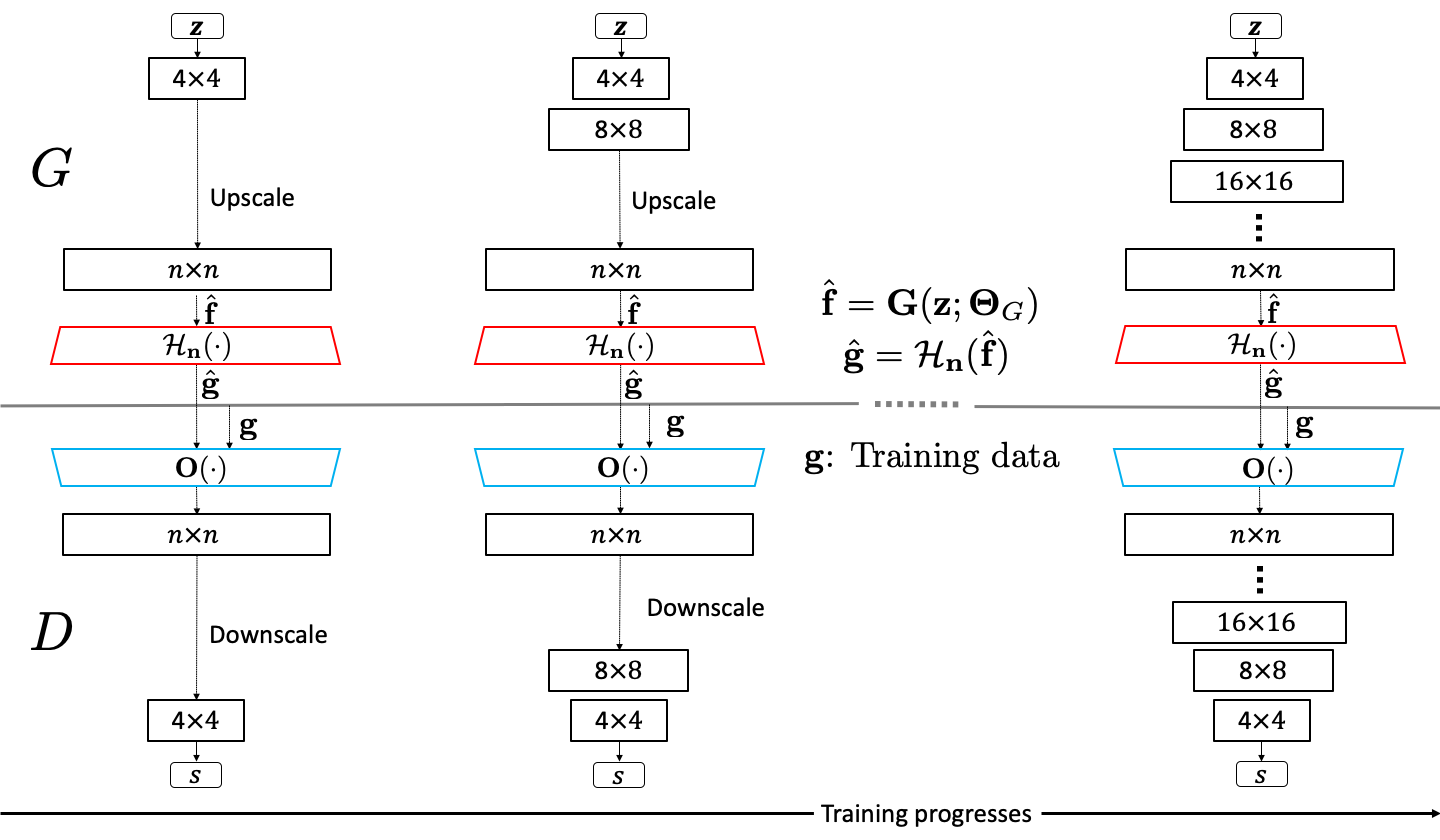}
  \caption{An illustration of ProAmGAN training. The training starts with low image resolution (e.g., $4\times 4$) and the image resolution is increased progressively by adding more layers to the generator and the discriminator. The discriminator is trained to distinguish between the ground-truth and generated reconstructed objects.}
  \label{fig:arc_PAGAN}
  \end{figure}

The ProAmGAN progressively grows the generator to establish the SOM from its low-resolution version to full-resolution version.
As with the AmbientGANs, the imaging measurements are subsequently simulated by applying the measurement operator to the generator-produced objects.
However, imaging measurements acquired in most medical imaging systems are indirect representations of objects to-be-imaged (e.g., Radon transform data, k-space data).
In such cases, 
the low-resolution version of the measured image data
and the low-resolution version of the objects may not be simply related
because they reside in generally different Hilbert spaces.
Accordingly, in these cases, the progressive growing strategy cannot be directly applied 
because the generator in the original ProGAN 
produces images that reside in the same Hilbert space as the training data employed by the discriminator. 
To address this issue, in addition to including the measurement operator as with the AmbientGAN training strategy, an image reconstruction operator
 $\mathcal{O}$: $\mathbb{R}^{M} \rightarrow \mathbb{R}^N$ 
 is included in the proposed ProAmGAN training strategy. 
 In this way, 
 the generator can be trained to produce images that reside in the same Hilbert space as the images employed by the discriminator
 and the progressive growing strategy can be subsequently employed.
The ProAmGAN training strategy is illustrated in Fig.~\ref{fig:arc_PAGAN}.
 
Given a training dataset that comprises measured data $\vec{g}$,
 a set of reconstructed objects $\vec{f}_{recon}$
 is computed by applying
 the operator $\mathcal{O}$ to the measured data $\vec{g}$:  $\vec{f}_{recon} = \mathcal{O}(\vec{g}) \equiv \mathcal{O}(\mathcal{H}_\vec{n}(\vec{f}))$.
 Denote the reconstructed object corresponding to the generator-produced measured data $\hat{\vec{g}}$  as $\hat{\vec{f}}_{recon}$: $\hat{\vec{f}}_{recon} = \mathcal{O}(\hat{\vec{g}}) \equiv \mathcal{O}\Big(\mathcal{H}_\vec{n}\big(\mathbf{G}(\mathbf{z}; \mathbf{\Theta}_{G}) \big) \Big)$. 
The discriminator in the ProAmGAN is trained to distinguish between $\hat{\vec{f}}_{recon}$ and $\vec{f}_{recon}$,
and the generator is trained to generate objects $\hat{\vec{f}} = \mathbf{G}(\mathbf{z}; \mathbf{\Theta}_{G})$ such that the corresponding reconstructed objects $\hat{\vec{f}}_{recon}$ are
indistinguishable from the reconstructed objects ${\vec{f}}_{recon}$ that were reconstructed from the provided measurement data (i.e., training data).
As with the AmbientGAN, when the distribution of objects $\mathrm{pr}(\vec{f})$ uniquely induces the distribution of reconstructed objects $\mathrm{pr}(\vec{f}_{recon})$, and the ProAmGAN achieves the global optimal at the final full-resolution stage, 
the trained generator can be employed to produce object samples drawn from the distribution $\mathrm{pr}(\vec{f})$.
In special cases where the imaging operator $\mathbf{H}$ is full-rank and the measurement noise $\vec{n} = \vec{0}$,
 ProAmGANs reduce to original ProGANs that are directly trained on objects.

\section{Numerical studies}
\label{sec:num}

Computer-simulation studies were conducted to demonstrate the ability of the proposed ProAmGAN to establish realistic SOMs from imaging measurements corresponding to different
 stylized imaging modalities.
Details regarding the design of the computer-simulation studies are provided below.

\subsection{Idealized direct imaging system}
\label{subsec:num_X}

An idealized direct imaging system that acquired chest radiographs, modeled as: $\vec{g} = \vec{f} + \vec{n}$, was considered first.
By design, it was assumed that the measurement noise was the only source
of image degradation. 
The motivation for this study was to demonstrate the ability of the ProAmGAN to learn an SOM from noisy images.

An NIH database of clinical chest X-ray images~\cite{wang2017chestx} was employed 
to serve as ground truth objects $\vec{f}$.
Three thousand images were selected from this dataset.
These images  
were centrally cropped and resized to the dimension of $512\times 512$
and were normalized to the range between 0 and 1.
A collection of 3000 simulated measured images $\vec{g}$ were produced
by adding independent and identically distributed (i.i.d.) Gaussian noise with zero mean and the standard deviation of $2\%$ to the
collection of objects  $\mathbf{f}$.
An example of the objects and  the corresponding noisy imaging measurement are
 shown in Fig. S. 7 in the Supplementary file.

From the ensemble of simulated measured data, with the knowledge of the measurement noise model, a ProAmGAN was trained to establish a SOM
that characterizes the  distribution of objects $\mathbf{f}$. The architecture of the generator and the discriminator employed in the ProAmGAN is described in Table S. 1 in the Supplementary file. {Because the idealized planar X-ray imaging system acquires direct representations of objects (i.e., $\mathbb{V}=\mathbb{U}$), the reconstruction operator $\mathcal{O}(\cdot)$ was set to be the identity operator in the ProAmGAN training process.}

{For comparison, by use of the same ensemble of simulated measured images $\mathbf{g}$, a ProGAN was trained.
In this case, the generator was trained to learn the distribution of measured images $\mathbf{g}$ 
themselves, which  are contaminated by measurement noise,  instead of learning 
the distribution of objects $\mathbf{f}$ (i.e., the SOM).
 The ProGAN employed a generator and discriminator with the same architectures as those employed in the ProAmGAN.}
 
The Fr\'{e}chet Inception Distance (FID)~\cite{heusel2017gans} score, 
a widely employed metric to evaluate the performance of generative models, 
was computed to evaluate the performance of the original ProGAN and the proposed ProAmGAN. 
The FID score quantifies the distance between the features extracted by the Inception-v3 network~\cite{szegedy2016rethinking} from the ground-truth (``real") and generated objects (``fake"). 
Lower FID score indicates better quality and diversity of the generated objects. 
The FID scores were computed by use of 3000 ground-truth objects, 3000 ProGAN-generated objects and 3000 ProAmGAN-generated objects.

The structural similarity index (SSIM)~\cite{wang2004image} is a figure-of-merit describing the similarity of two digital images.
As another form of evaluation, SSIM values were computed for different pairs of images.
First, 
SSIM values were computed from 500,000 random pairs of ground truth objects.
Next,  SSIM values were computed
from 500,000 random pairs of
 ProAmGAN-generated and ground truth objects.
Finally, as a comparison, 
SSIM values were computed
from 500,000 random pairs of
 ProGAN-generated and ground truth objects.
From these three collections of SSIM values, three histograms were formed.
The overlap area between any two of the histograms (i.e., empirical PDFs)
 and the two-sample Kolmogorov-Smirnov (KS) test statistics~\cite{young1977proof} were computed.

\subsection{Computed tomographic imaging system} 
A stylized tomographic imaging system was investigated next. 
This imaging system was described as: $\vec{g} = \mathcal{R}\vec{f} + \vec{n}$, 
where $\mathcal{R}$ denotes a 2D discrete Radon transform~\cite{kak2002principles} that maps a 2D object $\vec{f}$
 to a sinogram. The angular scanning range was 180 degrees and tomographic views were evenly
spaced  with a 1 degree angular step.

An NIH-sponsored database of clinical chest CT images~\cite{yan2018deeplesion} was employed to serve as ground truth objects $\vec{f}$. Three thousand images of dimension of $512\times 512$ were selected from this dataset and were normalized to the range between 0 and 1. 
A collection of 3000 measured data $\vec{g}$ were simulated by acting $\mathcal{R}$ on each object and adding i.i.d. Gaussian noise with a standard deviation of $10\%$. 
An example of the objects and the corresponding measured imaging data are shown in Fig. S. 8 in the Supplementary file.
  
From the collection of measured data $\vec{g}$, 
a set of reconstructed objects  $\vec{f}_{recon}$ was generated by
use of a filtered back-projection (FBP) reconstruction algorithm that employed a Ram-Lak filter.
With the knowledge of the imaging operator and the measurement noise model,
a ProAmGAN was subsequently trained 
by use of the reconstructed objects. 
 The ProAmGAN employed the generator and discriminator with the architectures described in Table S. \rom{1} (a) in the Supplementary file. 
 In the ProAmGAN training process, the Radon transform $\mathcal{R}$ and the FBP operator were applied to the generated objects as discussed in Sec. \ref{sec:ProAmGAN}.
 
{As a comparison,
a ProGAN was trained by use of reconstructed objects $\vec{f}_{recon}$.
The generator in the ProGAN was trained to learn the distribution of $\vec{f}_{recon}$ instead of learning the distribution of $\vec{f}$. The ProGAN employed a generator and discriminator with the same architectures as those employed in the ProAmGAN.
The FID scores and empirical PDFs of SSIM values were computed as described in Sec. \ref{subsec:num_X}.}

\subsection{MR imaging system with complete k-space data}
\label{ssec:MR-full-k-space}
A stylized MR imaging system that acquires fully-sampled k-space data was investigated. 
This imaging system was described as: $\vec{g} = \mathcal{F}(\vec{f}) + \vec{n}$, 
where $\mathcal{F}$ denotes a 2D discrete Fourier transform (DFT).
A database of clinical brain MR images~\cite{brain_mri} were employed to serve as ground truth objects $\vec{f}$. Three thousand images having the dimension of $512\times 512$ were selected from this dataset and were normalized to the range between 0 and 1. A collection of 3000 measured image data $\vec{g}$ were simulated by computing the 2D DFT of the objects and adding i.i.d. zero mean Gaussian noise 
with a standard deviation of 10 to both the real and imaginary components.
An example of the objects and the corresponding magnitude of the measured k-space data are shown in Fig. S. 9 in the Supplementary file.

From the ensemble of measured images, 
an ensemble of reconstructed images  $\vec{f}_{recon}$ was generated by
acting a 2D inverse discrete Fourier transform (IDFT) to each measured image data $\vec{g}$.
A ProAmGAN was subsequently trained to establish a SOM that characterizes the distribution of objects $\vec{f}$ 
by use of the ensemble of reconstructed images $\vec{f}_{recon}$. 
 The ProAmGAN employed a generator and discriminator with architectures described in Table S. \rom{1} (a) in the Supplementary file. 
 In the training process, the 2D DFT and IDFT were applied to the generator-produced objects as discussed in Sec. \ref{sec:ProAmGAN}.

{For comparison, a ProGAN was trained by use of reconstructed images $\vec{f}_{recon}$.
 The ProGAN employed a generator and discriminator with the same architectures as those employed in the ProAmGAN.
The FID score and empirical PDFs of SSIM values were also computed as described in Sec. \ref{subsec:num_X}.}

\subsection{MR imaging system with under-sampled k-space data} 
\label{ssec:MR_sampling}

MR imaging systems sometimes acquire under-sampled k-space data to accelerate the data-acquisition process.
In such cases, the imaging operator $\mathbf{H}$ has a non-trivial null space 
and only the measurement component $\vec{f}_{meas} = \mathbf{H}^{\dagger}\mathbf{H}\vec{f}$
can be observed through the imaging system.
Here, $\mathbf{H}^{\dagger}$ denotes the Moore-Penrose pseudo-inverse of $\mathbf{H}$ and can be computed by applying 
a 2D IDFT to the zero-filled k-space data.
In this study, the impact of k-space under-sampling on images produced by the ProAmGAN was investigated.

\begin{figure}[H]
   \centering
 \includegraphics[width=\linewidth]{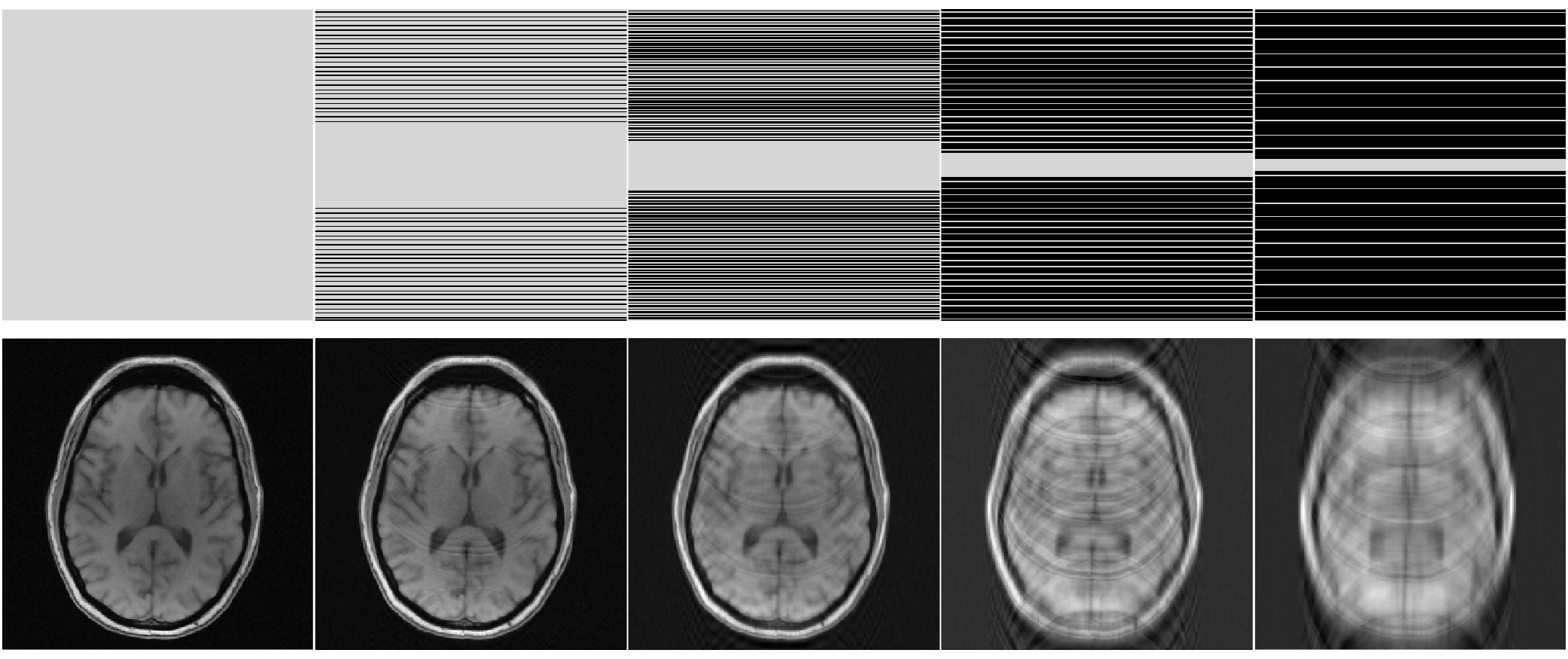}
 \vspace{-0.5cm}
 \caption{Top: k-space sampling patterns corresponding to different sampling ratios of ${1}/{1}$, ${4}/{5}$, ${1}/{2}$, ${1}/{4}$, and ${1}/{8}$ from left to right; 
Bottom: images reconstructed by use of $\mathbf{H}^{\dagger}$ corresponding to the k-space sampling patterns in the top row.}
 \label{fig:fastMRI}
  \vspace{-0.35cm}
\end{figure}

Clinical brain MR images contained in the NYU fastMRI Initiative database~\cite{zbontar2018fastmri} (\url{https:// fastmri.med.nyu.edu/})  were employed to serve as ground truth objects $\mathbf{f}$.
Three thousand images having dimension of $320\times 320$ were selected from this database for use in this study. 
These images were resized to the dimension of $256\times 256$ and were normalized to the range between 0 and 1.
Five data-acquisition designs corresponding to different k-space sampling ratios were considered: ${1}/{1}$, ${4}/{5}$, ${1}/{2}$, ${1}/{4}$, and ${1}/{8}$.
Here, the k-space sampling ratio was defined as the ratio of the number of sampled k-space components to the number of complete k-space components.
The sampling patterns are illustrated in the top row of Fig. \ref{fig:fastMRI}.
For each considered design, a collection of 3000 measured data $\vec{g}$ were simulated by computing and sampling the k-space data and adding i.i.d. zero mean Gaussian noise with a standard deviation of 2 to both the real and imaginary components.

For each data-acquisition design,
reconstructed objects $\vec{f}_{recon}$ were produced by acting the pseudo-inverse operator $\mathbf{H}^{\dagger}$ on the given measured image data $\vec{g}$.
Examples of reconstructed images using pseudo-inverse method corresponding to the considered sampling patterns are shown in the bottom row of Fig. \ref{fig:fastMRI}.
A ProAmGAN was subsequently trained to establish a SOM for each data-acquisition design.
The architecture of the generator and the discriminator employed in the ProAmGAN is described in Table S. \rom{1} (b) in the Supplementary file. 
 In the training process, $\mathbf{H}$ and $\mathbf{H}^{\dagger}$ were applied to the generator-produced objects as discussed in Sec. \ref{sec:ProAmGAN}.
 The FID score was computed by use of 3000 ground-truth objects $\vec{f}$ and 3000 ProAmGAN-generated objects $\hat{\vec{f}}$
 for each data-acquisition design. 
 Because only the measurement component $\vec{f}_{meas} = \mathbf{H}^{\dagger}\mathbf{H}\vec{f}$ can be measured by imaging systems, 
the ability of ProAmGANs to learn the variation in the measurement components was investigated.
Specifically, 
 the FID score was computed by use of the ground-truth measurement components $\vec{f}_{meas} = \mathbf{H}^{\dagger}\mathbf{H}\vec{f}$
 and ProAmGAN-generated measurement components $\hat{\vec{f}}_{meas}= \mathbf{H}^{\dagger}\mathbf{H}\hat{\vec{f}}$ for each data-acquisition design.

As a comparison, an original ProGAN was trained by use of the reconstructed objects $\vec{f}_{recon}$ for each data-acquisition design.
 The ProGAN employed the generator and the discriminator with the same architecture as those employed in the ProAmGAN.
The ProGAN-produced images were compared to the ProAmGAN-produced images.

\subsection{Task-based image quality assessment} 
\begin{figure}[H]
	\centering
	\includegraphics[width=\linewidth]{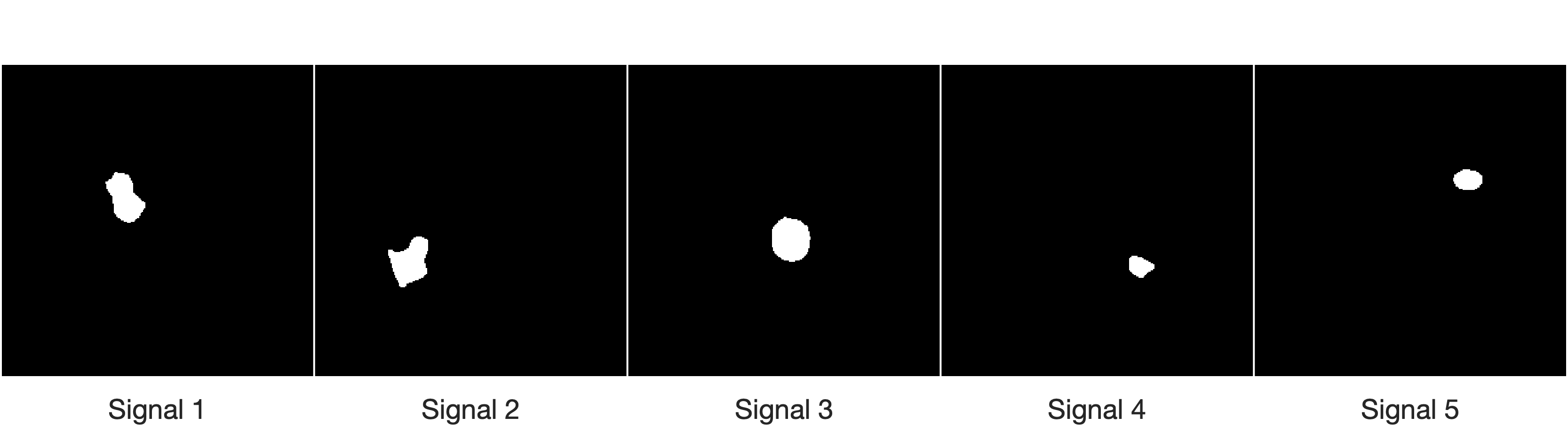}
	\caption{Five signals considered in the signal detection study.}
 \label{fig:signal}
\vspace{-0.15cm}
\end{figure}

In this study, the ProAmGAN-established SOMs corresponding to fastMRI brain objects were evaluated 
by use of objective measures of IQ.
Specifically, the ProAmGAN-established SOMs were evaluated 
by comparing task-specific image quality measures computed by use of generated objects to those computed by use of ground-truth objects.
A signal-known-exactly binary classification task was considered in which
an observer classifies noisy MR images as satisfying either a signal-absent hypothesis ($H_0$) or signal-present hypothesis ($H_1$).
The imaging processes under these two hypotheses can be described as:
\begin{subequations}
\label{eq:imgH_s}
\begin{align}
H_{0}:&\ \mathbf{g} = \mathbf{f} + \mathbf{n}, \\
H_{1}:&\  \mathbf{g} = \mathbf{f} + \mathbf{s} + \mathbf{n},
\end{align}
\end{subequations}
where $\mathbf{s}$ denotes a signal image and $\vec{n}$ is i.i.d. zero-mean Gaussian noise.
 Two different noise levels with standard deviations of $1\%$ and $5\%$, and five different signals  were considered. The considered signals are shown
 in Fig.~\ref{fig:signal}.

\vspace{-0.3cm}

Each considered signal detection task was performed on a region of interest (ROI) of dimension of $50\times 50$ pixels centered at the signal location.
The signal-to-noise ratio of the Hotelling observer (HO) test statistic $\text{SNR}_{HO}$ was employed as 
the figure-of-merit for assessing the image quality\cite{barrett2013foundations}:
\begin{equation}
\label{eq:snr}
  \text{SNR}_{HO} = \sqrt{{\vec{s}_{ROI}}^T\mathbf{K}^{-1}{\vec{s}_{ROI}}}, 
\end{equation}
where $\vec{s}_{ROI} \in \mathbb{R}^{2500\times 1}$ denotes the vectorized signal image in the ROI, 
and $\mathbf{K} \in \mathbb{R}^{2500\times 2500}$ denotes the covariance matrix corresponding to the ROIs in the noisy MR images.  
When computing 
$\text{SNR}_{HO}$, $\mathbf{K}^{-1}$ was calculated by use of a covariance matrix decomposition~\cite{barrett2013foundations}.
The values of $\text{SNR}_{HO}$  computed by use of 3000 ground truth objects and 3000 generated objects were compared.

\vspace{-0.2cm}
\subsection{Training details}
All ProAmGANs and ProGANs were trained by use of Tensorflow\cite{abadi2016tensorflow} by use of 4 NVIDIA Tesla V100 GPUs.
The Adam algorithm~\cite{kingma2014adam}, which is a stochastic gradient algorithm, was employed as the optimizer in the training process.
The ProAmGANs were implemented by modifying the ProGAN code ( \url{https://github.com/tkarras/progressive_growing_of_gans}) according to the proposed ProAmGAN architecture 
illustrated in Fig.~\ref{fig:arc_PAGAN}. Specifically, for each considered imaging system, the corresponding measurement operator and the reconstruction operator were applied to the generator-produced images, and the output images were subsequently employed by the discriminator.
The training of all ProAmGANs and ProGANs started with a resolution of $4\times 4$.
During the training process, the resolution was doubled by gradually
adding more layers to the generator and the discriminator until the final resolution was achieved.
More details regarding the progressive training details can be found in the literature \cite{karras2017progressive}.

\section{Results}
\label{sec:result}
\subsection{Visual assessments}
\begin{figure}[H]
	\centering
	\includegraphics[width=\linewidth]{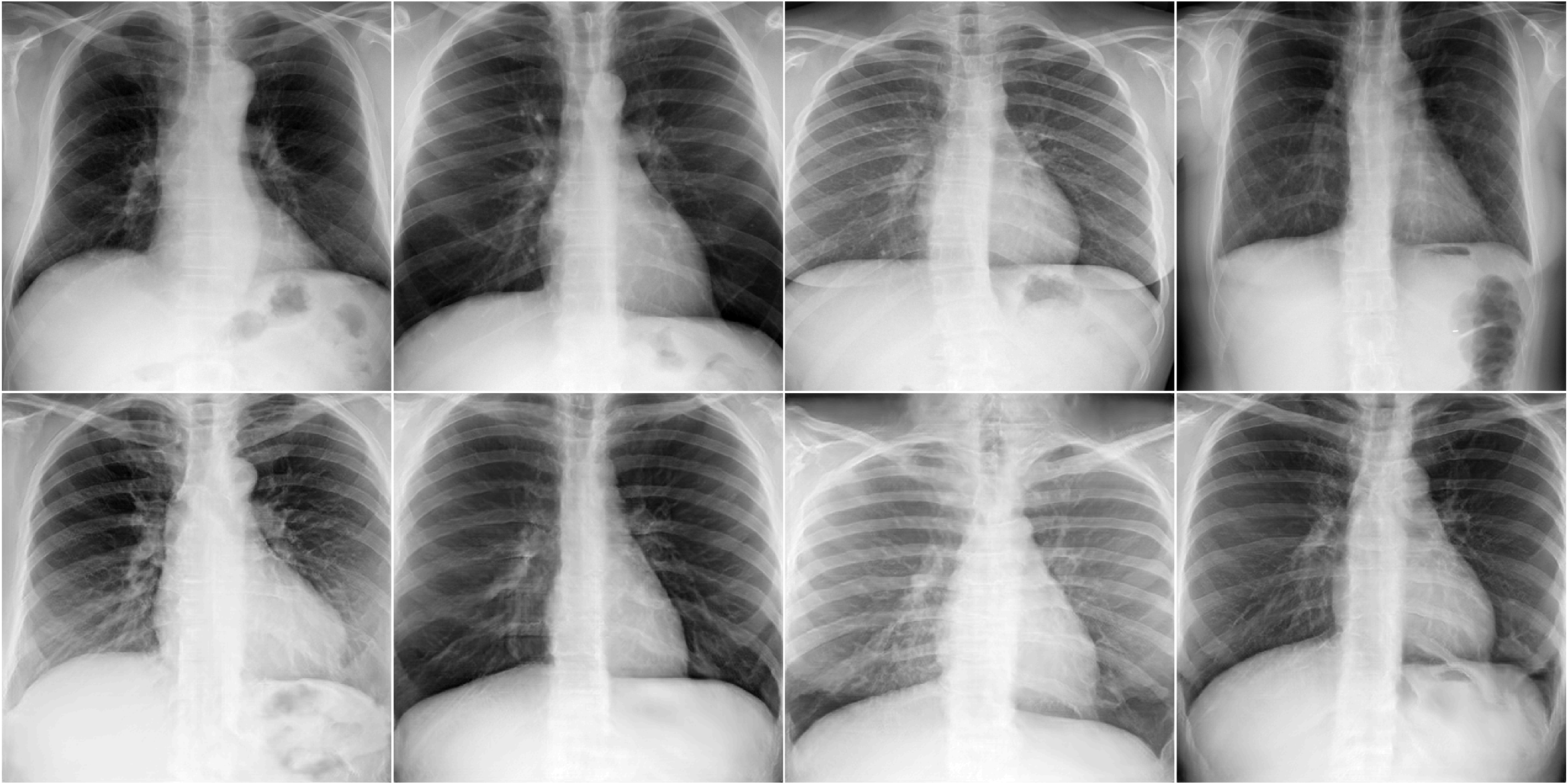}
\caption{Top: Ground-truth chest X-ray objects $\vec{f}$. Bottom: ProAmGAN-generated chest X-ray objects $\hat{\vec{f}}$.}
	\label{fig:x_fake}
 \vspace{-0.5cm}
\end{figure}

The ground-truth  (top row) and ProAmGAN-generated objects (bottom row)
corresponding to chest X-ray images
are shown in Fig.~\ref{fig:x_fake}.
The ProAmGAN-generated objects have similar visual appearances to the ground-truth ones.
Additional ProAmGAN-generated chest X-ray images are shown in Fig. S. 4 in the Supplementary file.
\vspace{-0.3cm}

\begin{figure}[H]
	\centering
	\includegraphics[width=0.85\linewidth]{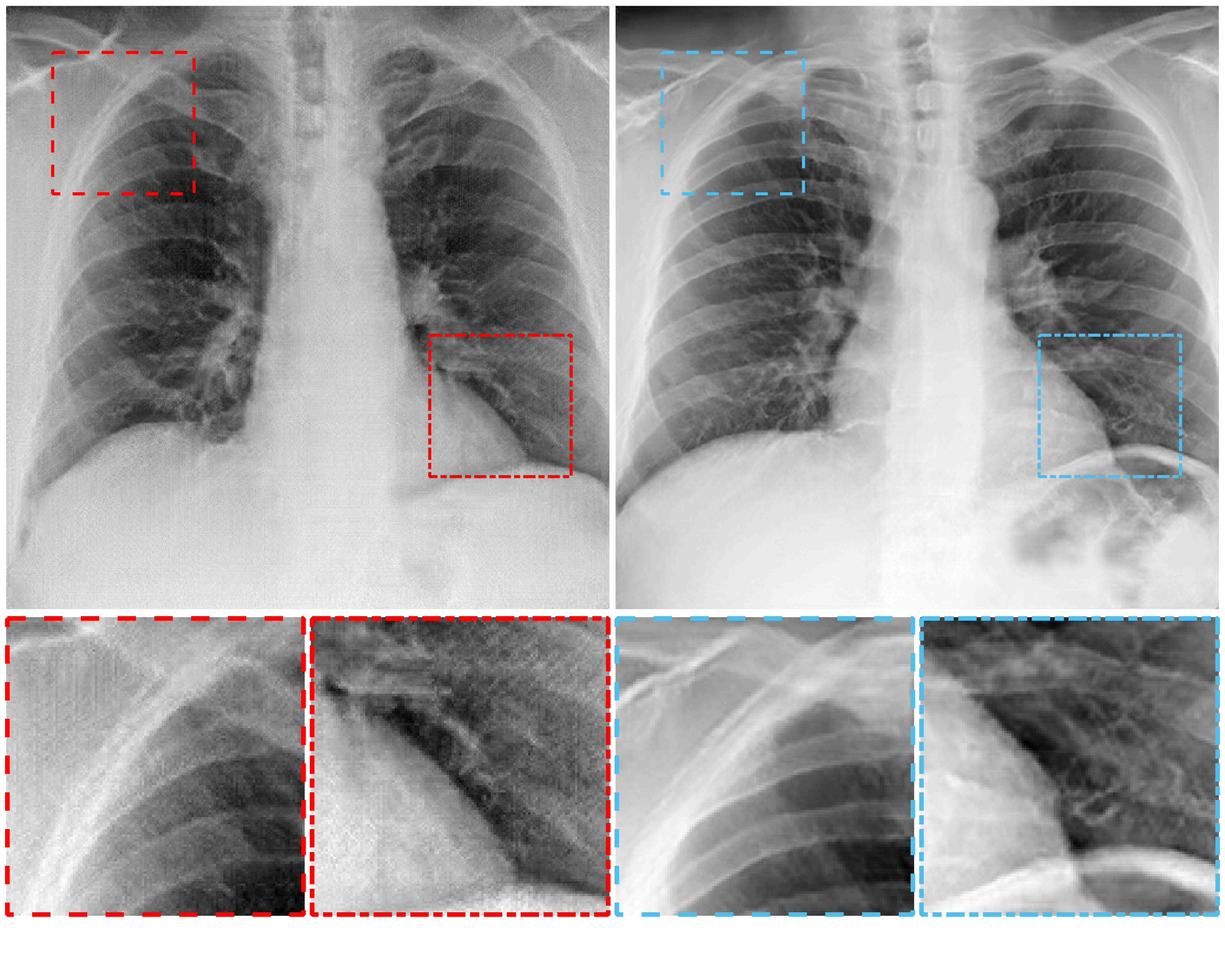}
	\vspace{-0.4cm}
	\caption{A ProGAN-generated (left panel) and ProAmGAN-generated (right panel) chest X-ray object.}
	\label{fig:progan}
\vspace{-0.3cm}
\end{figure}

A ProGAN-generated and ProAmGAN-generated objects are further compared in Fig.~\ref{fig:progan}.
It is clear that the ProAmGAN-produced chest X-ray image contains less noise than the one produced by the ProGAN.
This demonstrates the ability of the ProAmGAN to mitigate measurement noise when establishing SOMs.
\vspace{-0.3cm}

\begin{figure}[H]
	\centering
	\includegraphics[width=\linewidth]{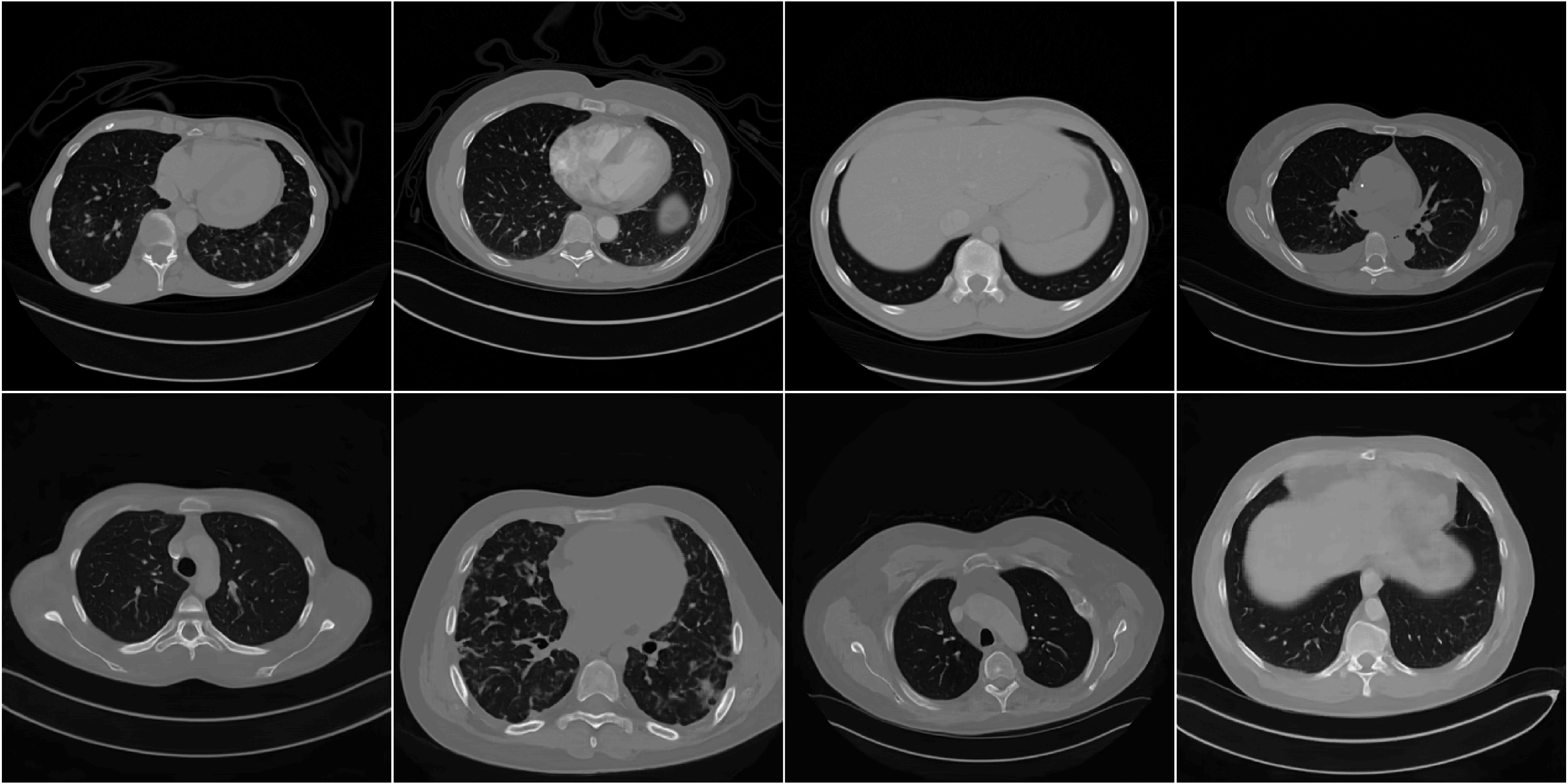}
	\vspace{-0.4cm}
	\caption{Top: Ground-truth chest CT objects $\vec{f}$. Bottom: ProAmGAN-generated chest CT objects $\hat{\vec{f}}$.}
	\label{fig:ct_fake}
\vspace{-0.25cm}
\end{figure}

\begin{figure}[H]
	\centering
	\includegraphics[width=\linewidth]{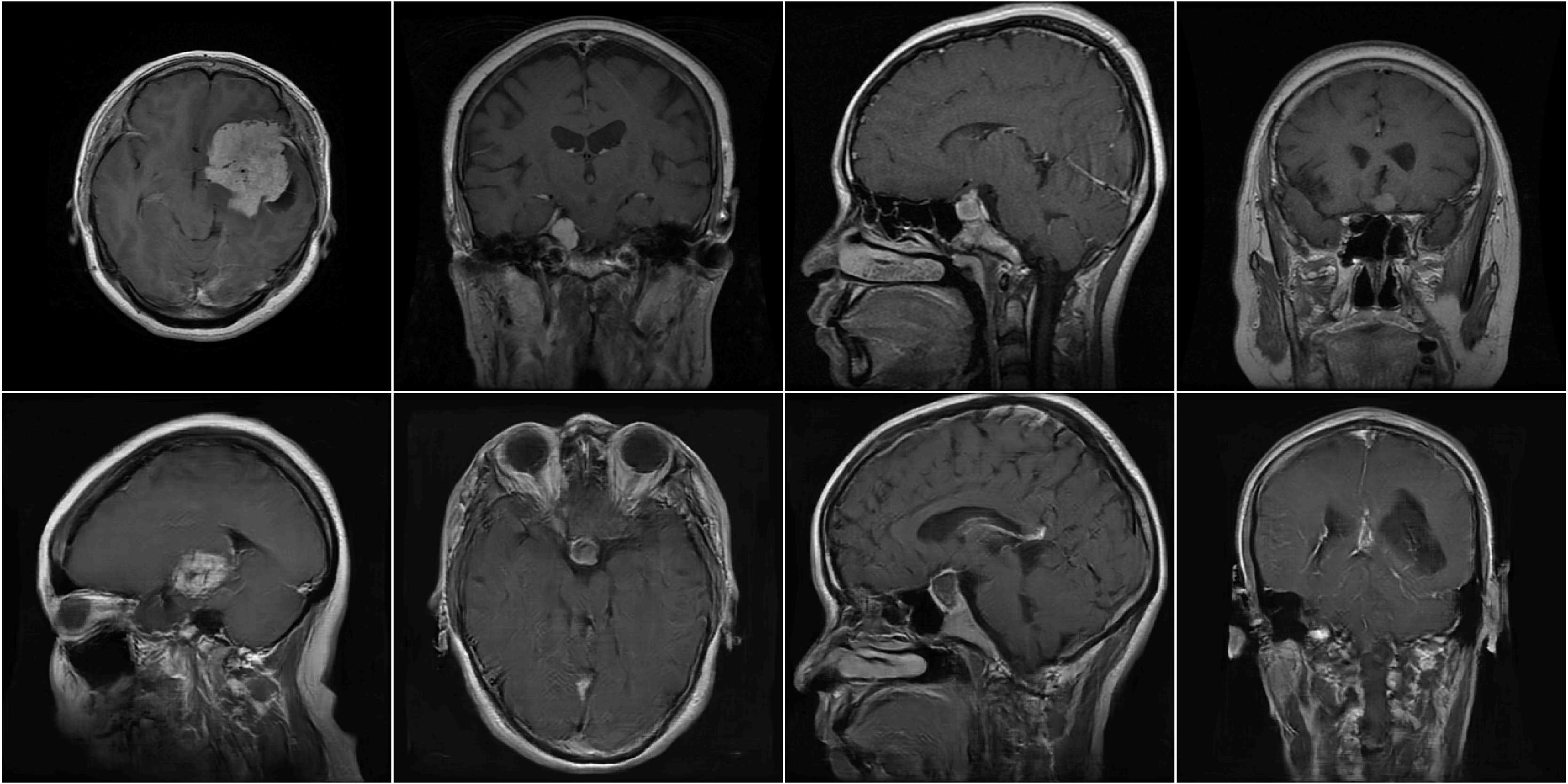}
	 \caption{Top: Ground-truth brain MR objects $\vec{f}$. Bottom: ProAmGAN-generated brain MR objects $\hat{\vec{f}}$.}
	\label{fig:mr_fake}
\vspace{-0.2cm}%
\end{figure}

The ground-truth (top row) and ProAmGAN-generated objects (bottom row) corresponding to chest CT and brain MR images are shown in Figs. \ref{fig:ct_fake} and \ref{fig:mr_fake}. 
The ProAmGAN-generated objects have similar visual appearances to ground-truth ones.
Additional ProAmGAN-generated chest CT images and brain MR images are shown in Figs. S. 5 and S. 6 in the Supplementary file.
\vspace{-0.25cm}
  
\begin{figure}[H]
	\centering
	\includegraphics[width=0.85\linewidth]{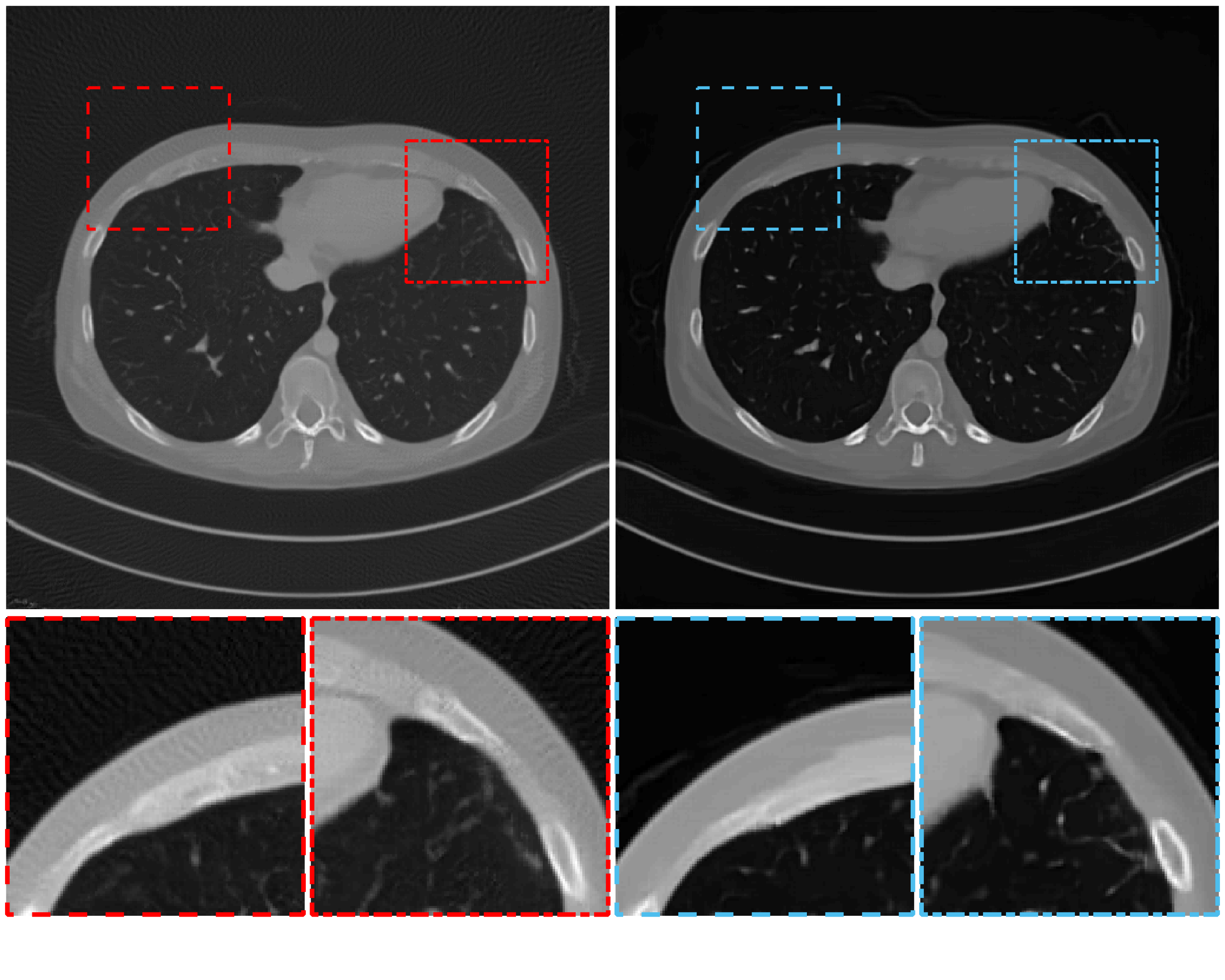}
	\vspace{-0.45cm}
	\caption{A ProGAN-generated (left panel) and ProAmGAN-generated (right panel) chest CT object.}
	\label{fig:progan_ct}
\vspace{-0.3cm}
\end{figure}

\begin{figure}[H]
	\centering
	\includegraphics[width=0.85\linewidth]{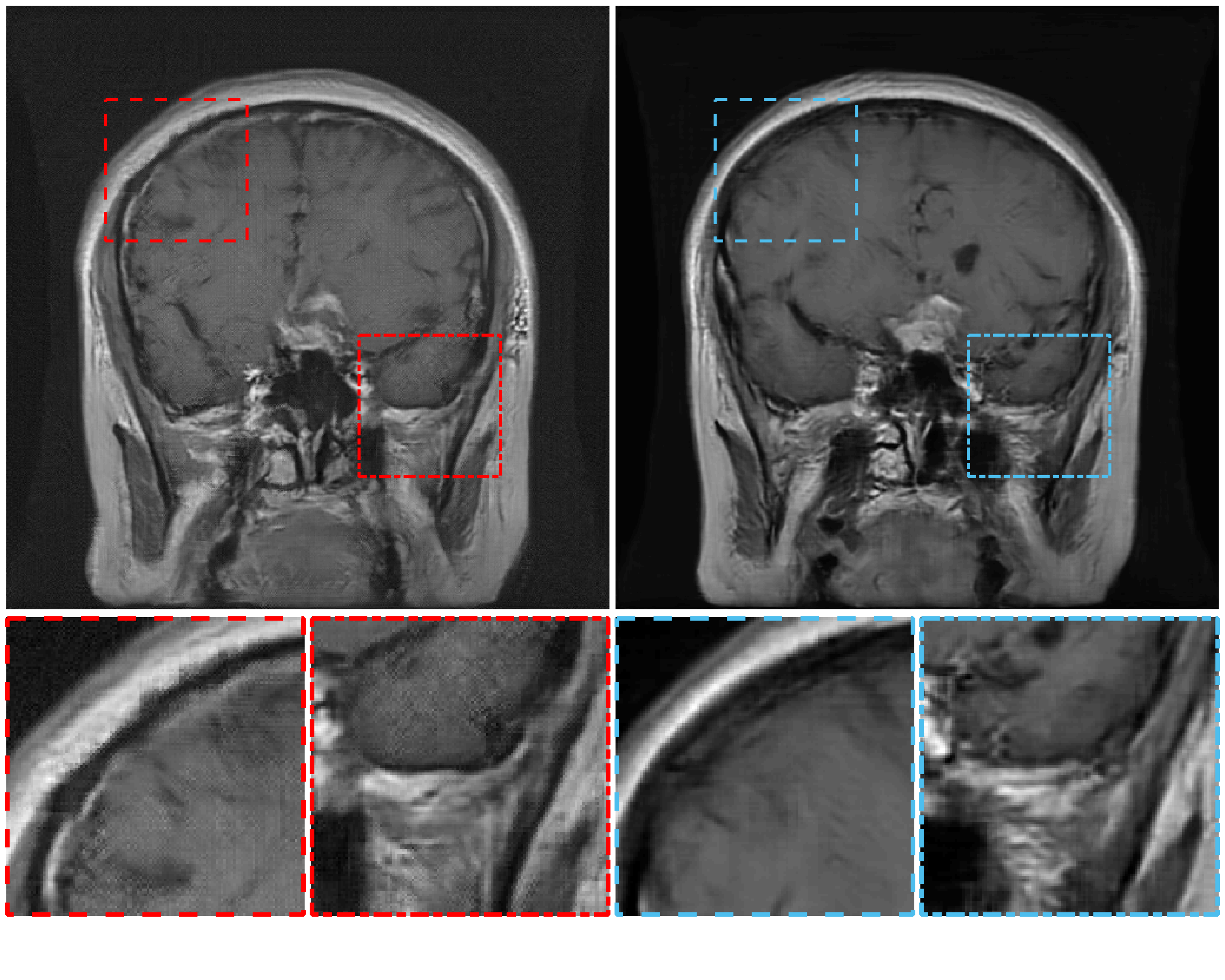}
	\vspace{-0.3cm}
	\caption{A ProGAN-generated (left panel) and ProAmGAN-generated (right panel) brain MR object.}
	\label{fig:mr_fake_compare}
\end{figure}

ProGAN-generated and ProAmGAN-generated objects are shown in more detail in Figs.~\ref{fig:progan_ct} and \ref{fig:mr_fake_compare}.
It is clear that the ProAmGAN-produced chest CT image in Fig.~\ref{fig:progan_ct} contains fewer artifacts than the one produced by the ProGAN.
This demonstrates the ability of the ProAmGAN to mitigate reconstruction artifacts when establishing SOMs.
\vspace{-0.35cm}

\vspace{-0.2cm}
  
The ProAmGAN-produced brain MR image in Fig. \ref{fig:mr_fake_compare} contains less noise than the one produced by the ProGAN.
This demonstrates the ability of the ProAmGAN to mitigate the noise in the reconstructed images when establishing SOMs.
\vspace{-0.2cm}

\subsection{Quantitative assessments}
The FID scores corresponding to ProGANs and ProAmGANs for the idealized direct imaging system, computed tomographic imaging system and MR imaging system with complete k-space data are shown in TABLE \ref{table:metrics}.
The ProAmGANs had smaller FID scores than the ProGANs,
which indicates that the ProAmGANs outperformed the ProGANs.
\begin{table}[H]
\begin{adjustbox}{width=\columnwidth,center}
\begin{tabular}{l|c|c|c|c|c|c}
\hline
\hline
                                                                       & \multicolumn{3}{c|}{ProGAN} & \multicolumn{3}{c}{ProAmGAN} \\ \cline{2-7} 
                                                                       & X-ray   & CT      & MRI     & X-ray    & CT       & MRI     \\ \hline
\begin{tabular}[c]{@{}l@{}}FID \\ score \   \end{tabular}                                                              & 65.583  & 62.385  & 47.247  & 28.798   & 30.616   & 41.637  \\ \hline
\begin{tabular}[c]{@{}l@{}}SSIM PDF \\ overlap area\end{tabular}       & 0.164   & 0.523   & 0.721   & 0.957    & 0.960    & 0.980   \\ \hline
\begin{tabular}[c]{@{}l@{}}Two-sample KS\\ test statistic\end{tabular} & 0.837   & 0.477   & 0.279   & 0.043    & 0.038    & 0.017   \\ \hline
\end{tabular}
\end{adjustbox}
\caption{FID and metrics that evaluate PDFs of SSIMs. Here, ``X-ray", ``CT'', and ``MRI"  correspond to the idealized direct imaging system, computed tomographic imaging system and MR imaging system with complete k-space data, respectively.}
\label{table:metrics}
\end{table}

\begin{figure}[H]
\centering
   \subfigure[Idealized direct imaging system]{\includegraphics[width=\linewidth]{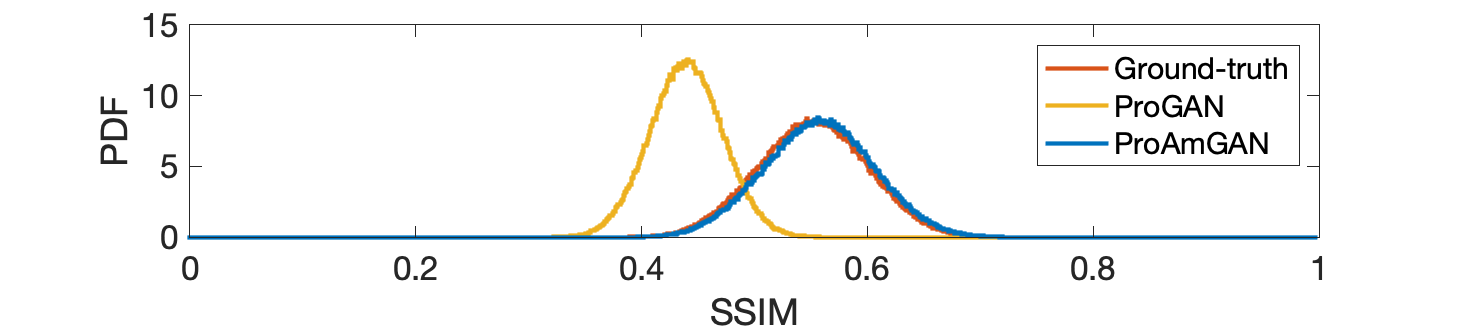}} \\[-0.5ex]
 \subfigure[Computed tomographic imaging system] {\includegraphics[width=\linewidth]{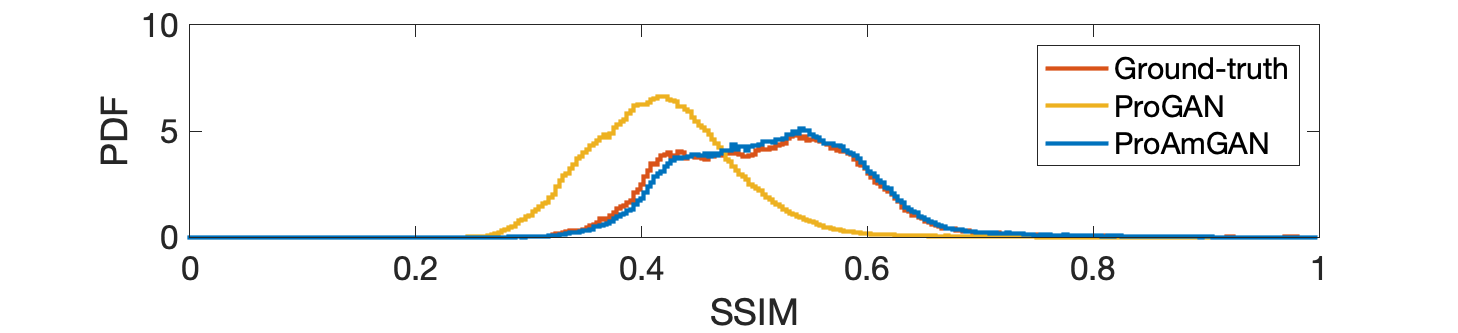}}\\[-0.5ex]
  \subfigure[MR imaging system with complete k-space data]{\includegraphics[width=\linewidth]{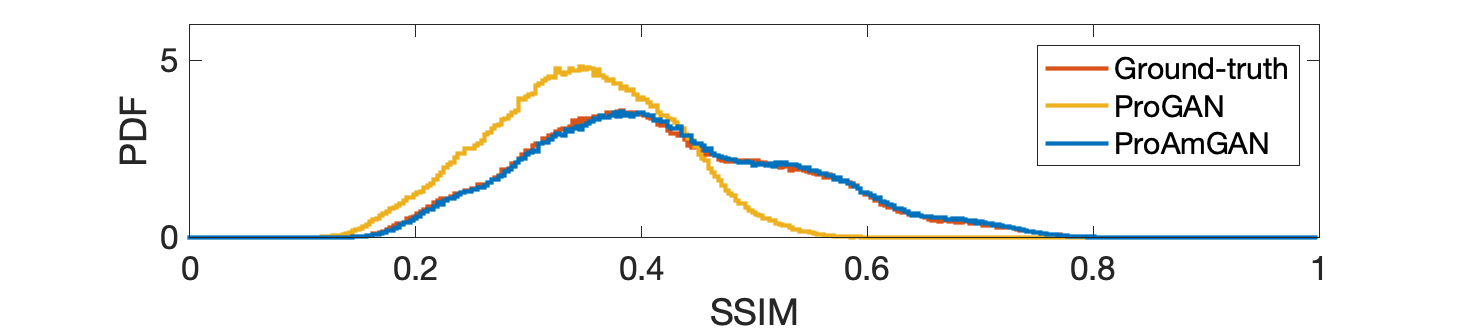}}\\[-0.5ex]
  \vspace{-0.1cm}
  \caption{Empirical PDFs of SSIMs corresponding to ground-truth image pairs (red curves), ground-truth and ProAmGAN-generated image pairs (blue curves), and ground-truth and ProGAN-generated image pairs (yellow curves).}
	\label{fig:ssim_x}
	\vspace{-0.2cm}
\end{figure}

The empirical PDFs of
SSIMs corresponding to the idealized direct imaging system, computed tomographic imaging system and MR imaging system with complete k-space data
are shown in Fig.~\ref{fig:ssim_x}, and the corresponding PDF overlap areas and two-sample KS test statistics are summarized in TABLE \ref{table:metrics}.
The PDFs of SSIMs corresponding to the ProAmGAN-generated and ground-truth objects largely overlap, while 
the one corresponding to the ProGAN-generated images had a significant discrepancy to the ground-truth PDF.
\vspace{-0.3cm}

\subsection{MR imaging system with under-sampled k-space data}
\begin{figure}[H]
   \centering
    \includegraphics[width=\linewidth]{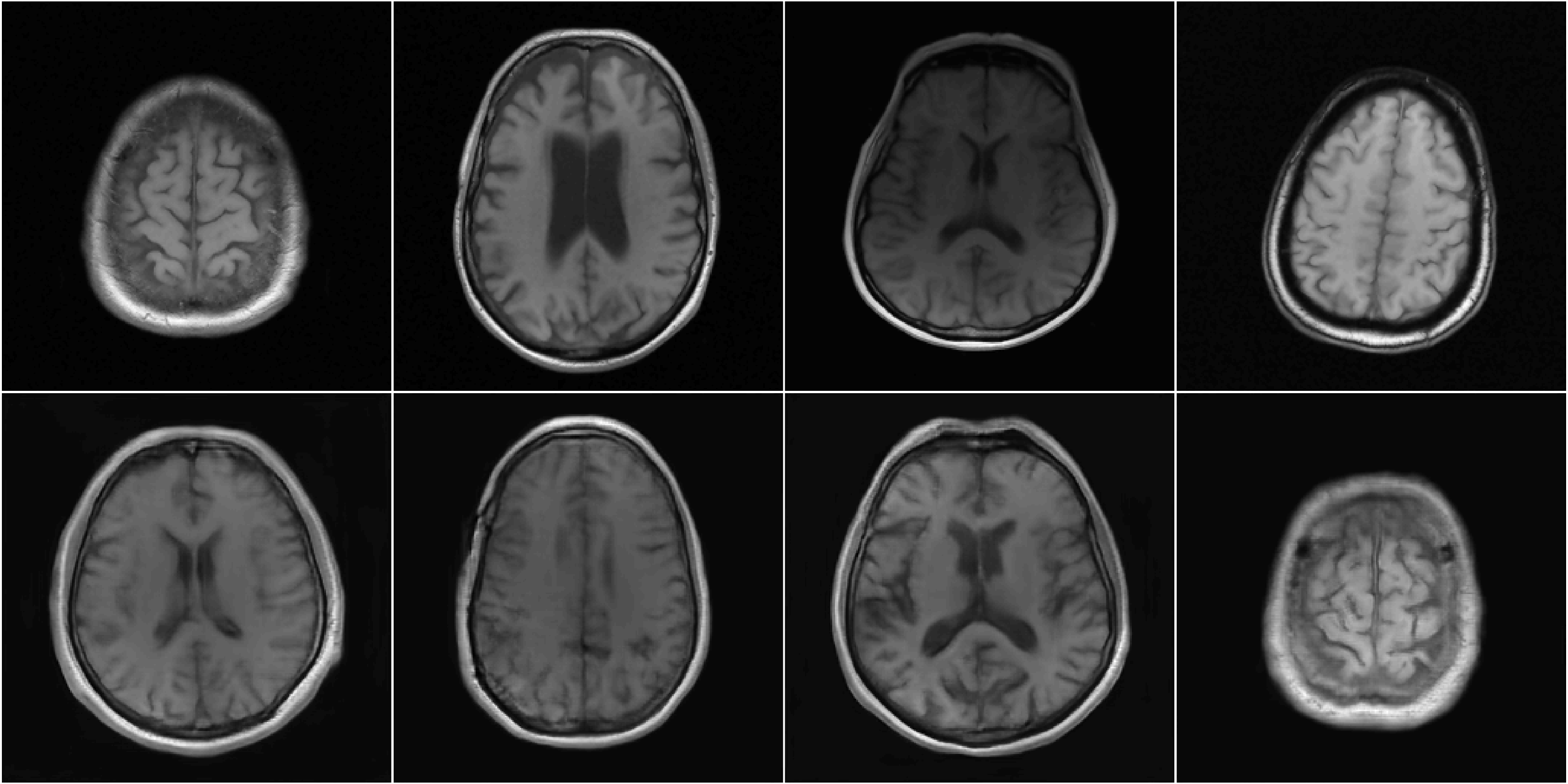}
 \caption{Top: Examples of ground-truth objects $\vec{f}$. Bottom: Examples of ProAmGAN-generated objects $\hat{\vec{f}}$ corresponding to the data-acquisition design with ${4}/{5}$  k-space sampling ratio.}
 \label{fig:fastMR}
 \vspace{-0.3cm}
\end{figure}

\vspace{-0.1cm}
The ground-truth (top row) objects and ProAmGAN-generated objects trained with ${4}/{5}$ k-space sampling ratio (bottom row) are shown in Fig. \ref{fig:fastMR}. 
The ProAmGAN-generated objects have similar visual appearances to the ground-truth objects.

\begin{figure}[H]
   \centering
 \includegraphics[width=\linewidth]{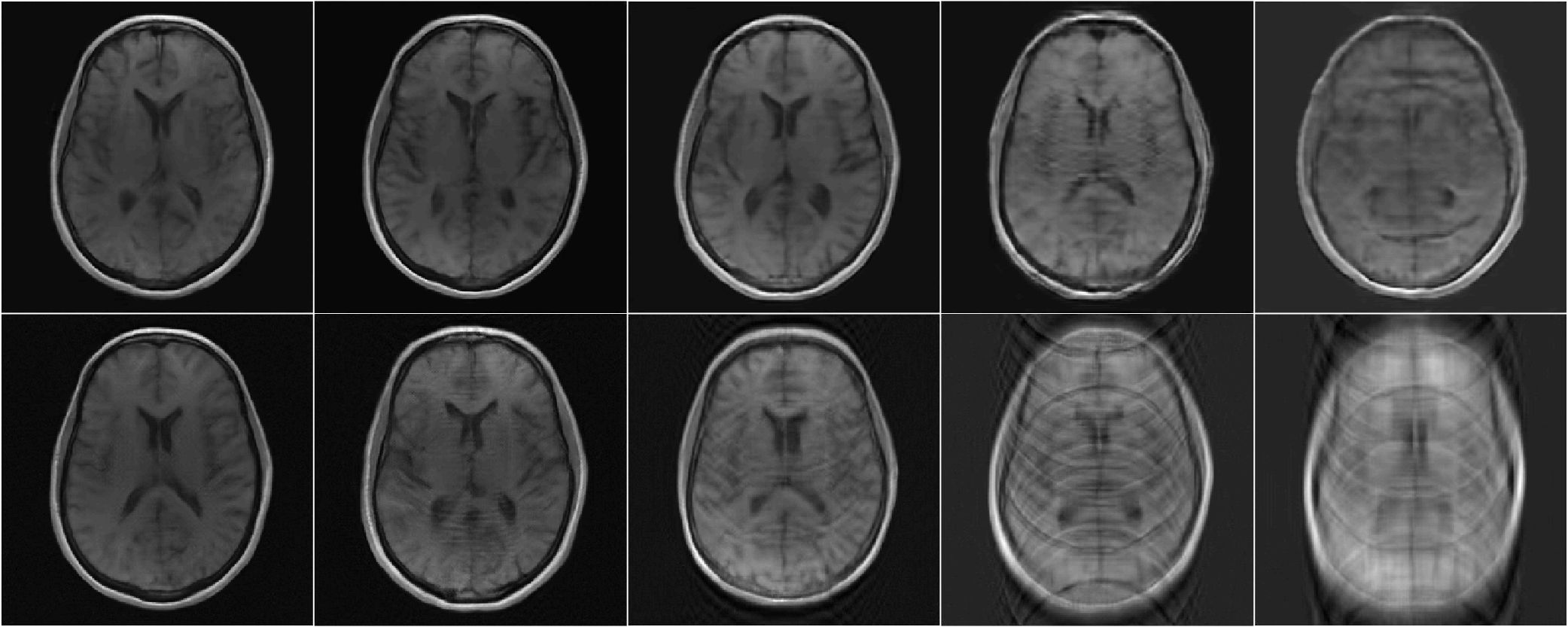}
 \caption{ProAmGAN-generated objects (top row) and ProGAN-generated objects (bottom row). From left to right, the ProGAN and ProAmGAN trained with the k-space sampling ratio of ${1}/{1}$, ${4}/{5}$, ${1}/{2}$, ${1}/{4}$, and ${1}/{8}$.}
 \label{fig:samplings}
 \vspace{-0.35cm}
\end{figure}

Objects produced by ProAmGANs and ProGANs trained with different data-acquisition designs are shown in Fig. \ref{fig:samplings}.
It was observed that the ProAmGAN-generated objects (top row)
are visually plausible for the k-space sampling ratios that range from ${1}/{2}$ to ${1}/{1}$, while
the noise and aliasing artifacts appear in the ProGAN-generated objects (bottom row).
\vspace{-0.2cm}

 \vspace{-0.25cm}

The FID corresponding to the objects $\vec{f}$ and that corresponding to the measurement components $\vec{f}_{meas}$ for each data-acquisition design are summarized in TABLE \ref{table:fids}.
It is observed that the FID between $\vec{f}$ and $\hat{\vec{f}}$  increased when the k-space sampling ratio decreased, while the FID between $\vec{f}_{meas}$ and $\hat{\vec{f}}_{meas}$ were not significantly changed. 
This indicates that the SOMs established by ProAmGANs can be affected by the null space of imaging operator, while the variation in the measurement components can be reliably learned.

\begin{table}[H]
\center
\begin{tabular}{l|c|c}
\hline\hline
             & FID for $\vec{f}$ & FID for $\vec{f}_{meas}$ \\ \hline
Full k-space & \multicolumn{2}{l}{\ \ \ \ \ \ \ \ 30.225} \\ \hline
4/5 k-space  & 38.510                         & 24.033                               \\ \hline
1/2 k-space  & 65.478                         & 20.383                               \\ \hline
1/4 k-space  & 105.607                        & 19.103                               \\ \hline
1/8 k-space  & 144.367                        & 20.122                               \\ \hline
\end{tabular}
\caption{FID scores corresponding to the objects and the measurement components. }
\label{table:fids}
\end{table}

\subsection{Task-based image quality assessment} 
\begin{figure}[H]
   \centering
 \includegraphics[width=0.95\linewidth]{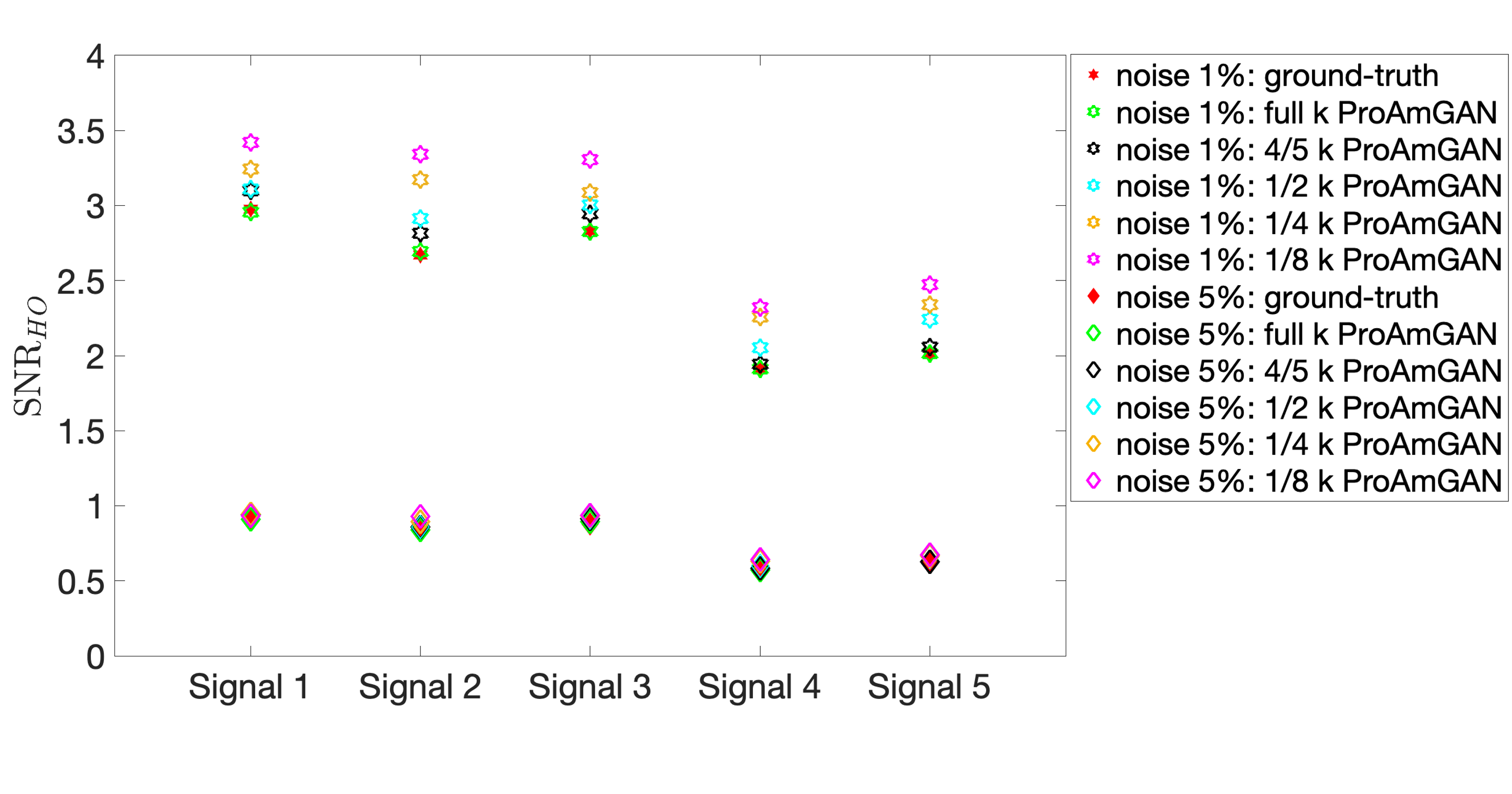}
 \vspace{-0.5cm}
 \caption{Hotelling observer performance corresponding to different tasks with  different signals, noise levels, and k-space sampling ratios.}
 \label{fig:HOs}
\end{figure}

The Hotelling observer performance was computed according to Eq. (\ref{eq:snr}) and is shown in Fig. \ref{fig:HOs}. It was observed that $SNR_{HO}$ has a positive bias when the ProAmGAN is trained with imaging systems that have large k-space missing ratios. This is because the ProAmGAN was not able to learn the complete object variation when the imaging system has a large null-space. 
When the noise level was increased, the object variation became relatively less important in terms of limiting the observer performance, and the positive bias of $SNR_{HO}$ subsequently became less significant.
This is consistent with the observation in reference \cite{kupinski2007bias}.
 \vspace{-0.3cm}

\section{Discussion and Conclusion}
\label{sec:conclusion}
Variation in  the objects to-be-imaged can significantly limit
the performance of an observer.
When conducting computer-simulation studies, this variation can be described by SOMs.
In this work,
a deep learning-based method that employed ProAmGANs was developed and investigated for establishing SOMs from measured image data.
The proposed ProAmGAN strategy incorporates the advanced progressive growing training procedure 
and therefore enables the AmbientGAN to be applied to realistically sized medical image data.
To demonstrate this, 
stylized numerical studies were conducted in which ProAmGANs were trained on different object ensembles corresponding to common medical imaging modalities.
Both visual examinations and quantitative analyses including task-specific validations indicate that the proposed ProAmGANs 
hold promise to establish realistic SOMs from imaging measurements.

In addition to objectively assessing imaging systems and data-acquisition designs,
the ProAmGAN-established SOMs can  be employed to regularize image reconstruction problems. 
Recent methods have been developed for regularizing image reconstruction problems based on GANs such as Compressed Sensing using Generative Models (CSGM)\cite{bora2017compressed} and image-adaptive GAN-based reconstruction methods (IAGAN)\cite{hussein2019image, bhadra2020medical}. 
These methods can be readily employed with the SOMs established by use of the proposed ProAmGANs.
ProAmGANs can also be used to produce clean reference images for training deep neural networks for 
solving other image-processing problems such as image denoising\cite{zhang2017beyond} and image super-resolution\cite{dong2014learning}.

It is desirable to establish three-dimensional (3D) object models.
A preliminary study developed a progressive-growing 3D GAN\cite{eklund2019feeding} and demonstrated its ability to generate 3D MR brain images with the dimension of $64\times 64\times 64$.
Our proposed method
can be readily extended to establish 3D object models
 by adopting such 3D GAN training strategies.
 Establishing a 3D version of the ProAmGAN will be explored in the future.
 
There remain additional topics for future investigation. 
It is critical to validate the learned SOMs for specific diagnostic tasks.
We have conducted preliminary 
task-specific validation studies by use of the Hotelling Observer \cite{barrett2013foundations, zhou2019learning_HO} and simple binary signal detection tasks.
It will be important to validate the learned SOMs for more complicated tasks by use of other observers such as the ideal observer\cite{zhou2018learning, zhou2019approximating, zhou2019learningIO, zhou2020markov} and anthropomorphic observers\cite{massanes2017evaluation}.
Finally, our proposed method can be readily employed with other GAN architectures such as 
the style-based generator architecture (StyleGAN)~\cite{karras2019style, karras2019analyzing} that can provide the additional ability to control certain features of generated-images and potentially can further improve the quality of generated-images. 

\bibliography{PAmbientGAN}{}

\begin{thebibliography}{10}
\providecommand{\url}[1]{#1}
\csname url@samestyle\endcsname
\providecommand{\newblock}{\relax}
\providecommand{\bibinfo}[2]{#2}
\providecommand{\BIBentrySTDinterwordspacing}{\spaceskip=0pt\relax}
\providecommand{\BIBentryALTinterwordstretchfactor}{4}
\providecommand{\BIBentryALTinterwordspacing}{\spaceskip=\fontdimen2\font plus
\BIBentryALTinterwordstretchfactor\fontdimen3\font minus
  \fontdimen4\font\relax}
\providecommand{\BIBforeignlanguage}[2]{{%
\expandafter\ifx\csname l@#1\endcsname\relax
\typeout{** WARNING: IEEEtran.bst: No hyphenation pattern has been}%
\typeout{** loaded for the language `#1'. Using the pattern for}%
\typeout{** the default language instead.}%
\else
\language=\csname l@#1\endcsname
\fi
#2}}
\providecommand{\BIBdecl}{\relax}
\BIBdecl

\bibitem{myers1993rayleigh}
K.~J. Myers, R.~F. Wagner, and K.~M. Hanson, ``Rayleigh task performance in
  tomographic reconstructions: {{\rm C}}omparison of human and machine
  performance,'' in \emph{Medical Imaging 1993: Image Processing}, vol.
  1898.\hskip 1em plus 0.5em minus 0.4em\relax International Society for Optics
  and Photonics, 1993, pp. 628--637.

\bibitem{wagner1985unified}
R.~F. Wagner and D.~G. Brown, ``Unified {SNR} analysis of medical imaging
  systems,'' \emph{Physics in Medicine \& Biology}, vol.~30, no.~6, p. 489,
  1985.

\bibitem{barrett1993model}
H.~H. Barrett, J.~Yao, J.~P. Rolland, and K.~J. Myers, ``Model observers for
  assessment of image quality,'' \emph{Proceedings of the National Academy of
  Sciences}, vol.~90, no.~21, pp. 9758--9765, 1993.

\bibitem{barrett2013foundations}
H.~H. Barrett and K.~J. Myers, \emph{Foundations of {I}mage {S}cience}.\hskip
  1em plus 0.5em minus 0.4em\relax John Wiley \&amp; Sons, 2013.

\bibitem{anastasio2010analysis}
M.~A. Anastasio, C.-Y. Chou, A.~M. Zysk, and J.~G. Brankov, ``Analysis of ideal
  observer signal detectability in phase-contrast imaging employing linear
  shift-invariant optical systems,'' \emph{JOSA A}, vol.~27, no.~12, pp.
  2648--2659, 2010.

\bibitem{rolland1992effect}
\BIBentryALTinterwordspacing
J.~P. Rolland and H.~H. Barrett, ``Effect of random background inhomogeneity on
  observer detection performance,'' \emph{J. Opt. Soc. Am. A}, vol.~9, no.~5,
  pp. 649--658, May 1992. [Online]. Available:
  \url{http://josaa.osa.org/abstract.cfm?URI=josaa-9-5-649}
\BIBentrySTDinterwordspacing

\bibitem{kupinski2003experimental}
M.~A. Kupinski, E.~Clarkson, J.~W. Hoppin, L.~Chen, and H.~H. Barrett,
  ``Experimental determination of object statistics from noisy images,''
  \emph{JOSA A}, vol.~20, no.~3, pp. 421--429, 2003.

\bibitem{bochud1999statistical}
F.~O. Bochud, C.~K. Abbey, and M.~P. Eckstein, ``Statistical texture synthesis
  of mammographic images with clustered lumpy backgrounds,'' \emph{Optics
  Express}, vol.~4, no.~1, pp. 33--43, 1999.

\bibitem{segars2008realistic}
W.~P. Segars, M.~Mahesh, T.~J. Beck, E.~C. Frey, and B.~M. Tsui, ``Realistic ct
  simulation using the 4d xcat phantom,'' \emph{Medical Physics}, vol.~35,
  no.~8, pp. 3800--3808, 2008.

\bibitem{segars2002study}
W.~P. Segars and B.~M. Tsui, ``Study of the efficacy of respiratory gating in
  myocardial spect using the new 4-d ncat phantom,'' \emph{IEEE Transactions on
  Nuclear Science}, vol.~49, no.~3, pp. 675--679, 2002.

\bibitem{xu2014exponential}
X.~G. Xu, ``An exponential growth of computational phantom research in
  radiation protection, imaging, and radiotherapy: a review of the fifty-year
  history,'' \emph{Physics in medicine and biology}, vol.~59, no.~18, p. R233,
  2014.

\bibitem{zankl2010gsf}
M.~Zankl and K.~Eckerman, ``The gsf voxel computational phantom family,''
  \emph{Handbook of Anatomical Models for Radiation Dosimetry}, pp. 65--85,
  2010.

\bibitem{collins1998design}
D.~L. Collins, A.~P. Zijdenbos, V.~Kollokian, J.~G. Sled, N.~J. Kabani, C.~J.
  Holmes, and A.~C. Evans, ``Design and construction of a realistic digital
  brain phantom,'' \emph{IEEE {T}ransactions on {M}edical {I}maging}, vol.~17,
  no.~3, pp. 463--468, 1998.

\bibitem{caon2004voxel}
M.~Caon, ``Voxel-based computational models of real human anatomy: a review,''
  \emph{Radiation and {E}nvironmental {B}iophysics}, vol.~42, no.~4, pp.
  229--235, 2004.

\bibitem{zu2005vip}
X.~G. Zu, ``The {VIP}-{M}an {M}odel-{A} {D}igital {H}uman {T}estbed for
  {R}adiation {S}iimulations,'' \emph{SAE {T}ransactions}, pp. 779--787, 2005.

\bibitem{li2009methodology}
C.~M. Li, W.~P. Segars, G.~D. Tourassi, J.~M. Boone, and J.~T. Dobbins~III,
  ``Methodology for generating a 3d computerized breast phantom from empirical
  data,'' \emph{Medical Physics}, vol.~36, no.~7, pp. 3122--3131, 2009.

\bibitem{cootes1995active}
T.~F. Cootes, C.~J. Taylor, D.~H. Cooper, and J.~Graham, ``Active shape
  models-their training and application,'' \emph{Computer vision and image
  understanding}, vol.~61, no.~1, pp. 38--59, 1995.

\bibitem{cootes2015active}
T.~Cootes, M.~Roberts, K.~Babalola, and C.~Taylor, ``{A}ctive {S}hape and
  {A}ppearance {M}odels,'' in \emph{Handbook of Biomedical Imaging}.\hskip 1em
  plus 0.5em minus 0.4em\relax Springer, 2015, pp. 105--122.

\bibitem{heimann2009statistical}
T.~Heimann and H.-P. Meinzer, ``Statistical shape models for 3d medical image
  segmentation: a review,'' \emph{Medical {I}mage {A}nalysis}, vol.~13, no.~4,
  pp. 543--563, 2009.

\bibitem{ferrari2010images}
V.~Ferrari, F.~Jurie, and C.~Schmid, ``From images to shape models for object
  detection,'' \emph{{I}nternational {J}ournal of {C}omputer {V}ision},
  vol.~87, no.~3, pp. 284--303, 2010.

\bibitem{shen2012detecting}
K.-k. Shen, J.~Fripp, F.~M{\'e}riaudeau, G.~Ch{\'e}telat, O.~Salvado,
  P.~Bourgeat, A.~D.~N. Initiative \emph{et~al.}, ``Detecting global and local
  hippocampal shape changes in alzheimer's disease using statistical shape
  models,'' \emph{Neuroimage}, vol.~59, no.~3, pp. 2155--2166, 2012.

\bibitem{gordillo2013state}
N.~Gordillo, E.~Montseny, and P.~Sobrevilla, ``State of the art survey on {MRI}
  brain tumor segmentation,'' \emph{Magnetic resonance imaging}, vol.~31,
  no.~8, pp. 1426--1438, 2013.

\bibitem{tomoshige2014conditional}
S.~Tomoshige, E.~Oost, A.~Shimizu, H.~Watanabe, and S.~Nawano, ``A conditional
  statistical shape model with integrated error estimation of the conditions;
  application to liver segmentation in non-contrast ct images,'' \emph{Medical
  {I}mage {A}nalysis}, vol.~18, no.~1, pp. 130--143, 2014.

\bibitem{ambellan2019automated}
F.~Ambellan, A.~Tack, M.~Ehlke, and S.~Zachow, ``Automated segmentation of knee
  bone and cartilage combining statistical shape knowledge and convolutional
  neural networks: Data from the osteoarthritis initiative,'' \emph{Medical
  {I}mage {A}nalysis}, vol.~52, pp. 109--118, 2019.

\bibitem{kupinski2005small}
M.~A. Kupinski and H.~H. Barrett, \emph{Small-animal SPECT imaging}.\hskip 1em
  plus 0.5em minus 0.4em\relax Springer, 2005, vol. 233.

\bibitem{goodfellow2014generative}
I.~Goodfellow, J.~Pouget-Abadie, M.~Mirza, B.~Xu, D.~Warde-Farley, S.~Ozair,
  A.~Courville, and Y.~Bengio, ``Generative adversarial nets,'' in
  \emph{Advances in Neural Information Processing Systems}, 2014, pp.
  2672--2680.

\bibitem{bora2018ambientgan}
A.~Bora, E.~Price, and A.~G. Dimakis, ``Ambientgan: {G}enerative models from
  lossy measurements,'' in \emph{International Conference on Learning
  Representations (ICLR)}, 2018.

\bibitem{zhou2019learning}
W.~Zhou, S.~Bhadra, F.~Brooks, and M.~A. Anastasio, ``Learning stochastic
  object model from noisy imaging measurements using ambientgans,'' in
  \emph{Medical Imaging 2019: Image Perception, Observer Performance, and
  Technology Assessment}, vol. 10952.\hskip 1em plus 0.5em minus 0.4em\relax
  International Society for Optics and Photonics, 2019, p. 109520M.

\bibitem{karras2017progressive}
T.~Karras, T.~Aila, S.~Laine, and J.~Lehtinen, ``Progressive {G}rowing of
  {GAN}s for improved quality, stability, and variation,'' \emph{arXiv preprint
  arXiv:1710.10196}, 2017.

\bibitem{goodfellow2016deeplearning}
I.~Goodfellow, Y.~Bengio, and A.~Courville, \emph{Deep Learning}.\hskip 1em
  plus 0.5em minus 0.4em\relax The MIT Press, 2016.

\bibitem{clarkson2002transformation}
E.~Clarkson, M.~A. Kupinski, and H.~H. Barrett, ``Transformation of
  characteristic functionals through imaging systems,'' \emph{Optics express},
  vol.~10, no.~13, pp. 536--539, 2002.

\bibitem{arjovsky2017towards}
\BIBentryALTinterwordspacing
M.~Arjovsky and L.~Bottou, ``Towards principled methods for training generative
  adversarial networks,'' in \emph{International Conference on Learning
  Representations (ICLR 2017)}, 2017. [Online]. Available:
  \url{http://leon.bottou.org/papers/arjovsky-bottou-2017}
\BIBentrySTDinterwordspacing

\bibitem{arora2017do}
\BIBentryALTinterwordspacing
S.~Arora and Y.~Zhang, ``Do {GANs} actually learn the distribution? an
  empirical study,'' \emph{CoRR}, vol. abs/1706.08224, 2017. [Online].
  Available: \url{http://arxiv.org/abs/1706.08224}
\BIBentrySTDinterwordspacing

\bibitem{denton2015deep}
\BIBentryALTinterwordspacing
E.~L. Denton, S.~Chintala, A.~Szlam, and R.~Fergus, ``Deep generative image
  models using a laplacian pyramid of adversarial networks,'' \emph{CoRR}, vol.
  abs/1506.05751, 2015. [Online]. Available:
  \url{http://arxiv.org/abs/1506.05751}
\BIBentrySTDinterwordspacing

\bibitem{radford2015unsupervised}
A.~{Radford}, L.~{Metz}, and S.~{Chintala}, ``{Unsupervised Representation
  Learning with Deep Convolutional Generative Adversarial Networks},''
  \emph{arXiv e-prints}, p. arXiv:1511.06434, Nov 2015.

\bibitem{salimans2016improved}
\BIBentryALTinterwordspacing
T.~Salimans, I.~Goodfellow, W.~Zaremba, V.~Cheung, A.~Radford, and X.~Chen,
  ``Improved techniques for training gans,'' in \emph{Proceedings of the 30th
  International Conference on Neural Information Processing Systems}, ser.
  NIPS'16.\hskip 1em plus 0.5em minus 0.4em\relax USA: Curran Associates Inc.,
  2016, pp. 2234--2242. [Online]. Available:
  \url{http://dl.acm.org/citation.cfm?id=3157096.3157346}
\BIBentrySTDinterwordspacing

\bibitem{li2019misGAN}
\BIBentryALTinterwordspacing
S.~C. Li, B.~Jiang, and B.~M. Marlin, ``Misgan: Learning from incomplete data
  with generative adversarial networks,'' \emph{CoRR}, vol. abs/1902.09599,
  2019. [Online]. Available: \url{http://arxiv.org/abs/1902.09599}
\BIBentrySTDinterwordspacing

\bibitem{shrivastava2015learning}
\BIBentryALTinterwordspacing
A.~Shrivastava, T.~Pfister, O.~Tuzel, J.~Susskind, W.~Wang, and R.~Webb,
  ``Learning from simulated and unsupervised images through adversarial
  training,'' \emph{CoRR}, vol. abs/1612.07828, 2016. [Online]. Available:
  \url{http://arxiv.org/abs/1612.07828}
\BIBentrySTDinterwordspacing

\bibitem{arjovsky2017wasserstein}
M.~{Arjovsky}, S.~{Chintala}, and L.~{Bottou}, ``{Wasserstein GAN},''
  \emph{arXiv e-prints}, p. arXiv:1701.07875, Jan 2017.

\bibitem{gulrajani2017improved}
\BIBentryALTinterwordspacing
I.~Gulrajani, F.~Ahmed, M.~Arjovsky, V.~Dumoulin, and A.~C. Courville,
  ``Improved training of wasserstein gans,'' \emph{CoRR}, vol. abs/1704.00028,
  2017. [Online]. Available: \url{http://arxiv.org/abs/1704.00028}
\BIBentrySTDinterwordspacing

\bibitem{brock2018large}
\BIBentryALTinterwordspacing
A.~Brock, J.~Donahue, and K.~Simonyan, ``Large scale {GAN} training for high
  fidelity natural image synthesis,'' \emph{CoRR}, vol. abs/1809.11096, 2018.
  [Online]. Available: \url{http://arxiv.org/abs/1809.11096}
\BIBentrySTDinterwordspacing

\bibitem{wang2017chestx}
X.~Wang, Y.~Peng, L.~Lu, Z.~Lu, M.~Bagheri, and R.~M. Summers, ``Chestx-ray8:
  Hospital-scale chest x-ray database and benchmarks on weakly-supervised
  classification and localization of common thorax diseases,'' in
  \emph{Proceedings of the IEEE {C}onference on {C}omputer {V}ision and
  {P}attern {R}ecognition}, 2017, pp. 2097--2106.

\bibitem{heusel2017gans}
M.~Heusel, H.~Ramsauer, T.~Unterthiner, B.~Nessler, and S.~Hochreiter, ``Gans
  trained by a two time-scale update rule converge to a local nash
  equilibrium,'' in \emph{Advances in Neural Information Processing Systems},
  2017, pp. 6626--6637.

\bibitem{szegedy2016rethinking}
C.~Szegedy, V.~Vanhoucke, S.~Ioffe, J.~Shlens, and Z.~Wojna, ``Rethinking the
  inception architecture for computer vision,'' in \emph{Proceedings of the
  IEEE {C}onference on {C}omputer {V}ision and {P}attern {R}ecognition}, 2016,
  pp. 2818--2826.

\bibitem{wang2004image}
Z.~Wang, A.~C. Bovik, H.~R. Sheikh, and E.~P. Simoncelli, ``Image quality
  assessment: from error visibility to structural similarity,'' \emph{IEEE
  {T}ransactions on {I}mage {P}rocessing}, vol.~13, no.~4, pp. 600--612, 2004.

\bibitem{young1977proof}
I.~T. Young, ``Proof without prejudice: use of the kolmogorov-smirnov test for
  the analysis of histograms from flow systems and other sources.''
  \emph{Journal of Histochemistry \& Cytochemistry}, vol.~25, no.~7, pp.
  935--941, 1977.

\bibitem{kak2002principles}
A.~C. Kak, M.~Slaney, and G.~Wang, ``Principles of computerized tomographic
  imaging,'' \emph{Medical Physics}, vol.~29, no.~1, pp. 107--107, 2002.

\bibitem{yan2018deeplesion}
K.~Yan, X.~Wang, L.~Lu, and R.~M. Summers, ``Deeplesion: automated mining of
  large-scale lesion annotations and universal lesion detection with deep
  learning,'' \emph{Journal of Medical Imaging}, vol.~5, no.~3, p. 036501,
  2018.

\bibitem{brain_mri}
\BIBentryALTinterwordspacing
J.~Cheng, ``Brain tumor dataset.'' [Online]. Available:
  \url{https://doi.org/10.6084/m9.figshare.1512427.v5}
\BIBentrySTDinterwordspacing

\bibitem{zbontar2018fastmri}
J.~Zbontar, F.~Knoll, A.~Sriram, M.~J. Muckley, M.~Bruno, A.~Defazio,
  M.~Parente, K.~J. Geras, J.~Katsnelson, H.~Chandarana \emph{et~al.},
  ``fastmri: An open dataset and benchmarks for accelerated mri,'' \emph{arXiv
  preprint arXiv:1811.08839}, 2018.

\bibitem{abadi2016tensorflow}
M.~Abadi, P.~Barham, J.~Chen, Z.~Chen, A.~Davis, J.~Dean, M.~Devin,
  S.~Ghemawat, G.~Irving, M.~Isard \emph{et~al.}, ``Tensorflow: a system for
  large-scale machine learning.'' in \emph{OSDI}, vol.~16, 2016, pp. 265--283.

\bibitem{kingma2014adam}
D.~P. Kingma and J.~Ba, ``Adam: {A} method for stochastic optimization,''
  \emph{arXiv preprint arXiv:1412.6980}, 2014.

\bibitem{kupinski2007bias}
M.~A. Kupinski, E.~Clarkson, and J.~Y. Hesterman, ``Bias in {H}otelling
  observer performance computed from finite data,'' in \emph{Medical Imaging
  2007: Image Perception, Observer Performance, and Technology Assessment},
  vol. 6515.\hskip 1em plus 0.5em minus 0.4em\relax International Society for
  Optics and Photonics, 2007, p. 65150S.

\bibitem{bora2017compressed}
A.~Bora, A.~Jalal, E.~Price, and A.~G. Dimakis, ``Compressed sensing using
  generative models,'' in \emph{Proceedings of the 34th International
  Conference on Machine Learning-Volume 70}.\hskip 1em plus 0.5em minus
  0.4em\relax JMLR. org, 2017, pp. 537--546.

\bibitem{hussein2019image}
S.~A. Hussein, T.~Tirer, and R.~Giryes, ``Image-adaptive gan based
  reconstruction,'' \emph{arXiv preprint arXiv:1906.05284}, 2019.

\bibitem{bhadra2020medical}
S.~Bhadra, W.~Zhou, and M.~A. Anastasio, ``Medical image reconstruction with
  image-adaptive priors learned by use of generative adversarial networks,'' in
  \emph{Medical Imaging 2020: Physics of Medical Imaging}, vol. 11312.\hskip
  1em plus 0.5em minus 0.4em\relax International Society for Optics and
  Photonics, 2020, p. 113120V.

\bibitem{zhang2017beyond}
K.~Zhang, W.~Zuo, Y.~Chen, D.~Meng, and L.~Zhang, ``Beyond a gaussian denoiser:
  Residual learning of deep cnn for image denoising,'' \emph{IEEE Transactions
  on Image Processing}, vol.~26, no.~7, pp. 3142--3155, 2017.

\bibitem{dong2014learning}
C.~Dong, C.~C. Loy, K.~He, and X.~Tang, ``Learning a deep convolutional network
  for image super-resolution,'' in \emph{European {C}onference on {C}omputer
  {V}ision}.\hskip 1em plus 0.5em minus 0.4em\relax Springer, 2014, pp.
  184--199.

\bibitem{eklund2019feeding}
A.~Eklund, ``Feeding the zombies: Synthesizing brain volumes using a 3d
  progressive growing gan,'' \emph{arXiv preprint arXiv:1912.05357}, 2019.

\bibitem{zhou2019learning_HO}
W.~Zhou, H.~Li, and M.~A. Anastasio, ``Learning the {H}otelling observer for
  {SKE} detection tasks by use of supervised learning methods,'' in
  \emph{Medical Imaging 2019: Image Perception, Observer Performance, and
  Technology Assessment}, vol. 10952.\hskip 1em plus 0.5em minus 0.4em\relax
  International Society for Optics and Photonics, 2019, p. 1095208.

\bibitem{zhou2018learning}
W.~Zhou and M.~A. Anastasio, ``Learning the {I}deal {O}bserver for {SKE}
  detection tasks by use of convolutional neural networks,'' in \emph{Medical
  Imaging 2018: Image Perception, Observer Performance, and Technology
  Assessment}, vol. 10577.\hskip 1em plus 0.5em minus 0.4em\relax International
  Society for Optics and Photonics, 2018, p. 1057719.

\bibitem{zhou2019approximating}
W.~Zhou, H.~Li, and M.~A. Anastasio, ``Approximating the {I}deal {O}bserver and
  {H}otelling {O}bserver for binary signal detection tasks by use of supervised
  learning methods,'' \emph{IEEE {T}ransactions on {M}edical {I}maging},
  vol.~38, no.~10, pp. 2456--2468, 2019.

\bibitem{zhou2019learningIO}
W.~Zhou and M.~A. Anastasio, ``Learning the ideal observer for joint detection
  and localization tasks by use of convolutional neural networks,'' in
  \emph{Medical Imaging 2019: Image Perception, Observer Performance, and
  Technology Assessment}, vol. 10952.\hskip 1em plus 0.5em minus 0.4em\relax
  International Society for Optics and Photonics, 2019, p. 1095209.

\bibitem{zhou2020markov}
W.~{Z}hou and M.~A. Anastasio, ``Markov-chain monte carlo approximation of the
  ideal observer using generative adversarial networks,'' in \emph{Medical
  Imaging 2020: Image Perception, Observer Performance, and Technology
  Assessment}, vol. 11316.\hskip 1em plus 0.5em minus 0.4em\relax International
  Society for Optics and Photonics, 2020, p. 113160D.

\bibitem{massanes2017evaluation}
F.~Massanes and J.~G. Brankov, ``Evaluation of cnn as anthropomorphic model
  observer,'' in \emph{Medical Imaging 2017: Image Perception, Observer
  Performance, and Technology Assessment}, vol. 10136.\hskip 1em plus 0.5em
  minus 0.4em\relax International Society for Optics and Photonics, 2017, p.
  101360Q.

\bibitem{karras2019style}
T.~Karras, S.~Laine, and T.~Aila, ``A style-based generator architecture for
  generative adversarial networks,'' in \emph{Proceedings of the IEEE
  Conference on Computer Vision and Pattern Recognition}, 2019, pp. 4401--4410.

\bibitem{karras2019analyzing}
T.~Karras, S.~Laine, M.~Aittala, J.~Hellsten, J.~Lehtinen, and T.~Aila,
  ``Analyzing and improving the image quality of stylegan,'' \emph{arXiv
  preprint arXiv:1912.04958}, 2019.

\end{thebibliography}


\begin{thebibliography}{1}
\providecommand{\url}[1]{#1}
\csname url@samestyle\endcsname
\providecommand{\newblock}{\relax}
\providecommand{\bibinfo}[2]{#2}
\providecommand{\BIBentrySTDinterwordspacing}{\spaceskip=0pt\relax}
\providecommand{\BIBentryALTinterwordstretchfactor}{4}
\providecommand{\BIBentryALTinterwordspacing}{\spaceskip=\fontdimen2\font plus
\BIBentryALTinterwordstretchfactor\fontdimen3\font minus
  \fontdimen4\font\relax}
\providecommand{\BIBforeignlanguage}[2]{{%
\expandafter\ifx\csname l@#1\endcsname\relax
\typeout{** WARNING: IEEEtran.bst: No hyphenation pattern has been}%
\typeout{** loaded for the language `#1'. Using the pattern for}%
\typeout{** the default language instead.}%
\else
\language=\csname l@#1\endcsname
\fi
#2}}
\providecommand{\BIBdecl}{\relax}
\BIBdecl

\bibitem{karras2017progressive}
T.~Karras, T.~Aila, S.~Laine, and J.~Lehtinen, ``Progressive {G}rowing of
  {GAN}s for improved quality, stability, and variation,'' \emph{arXiv preprint
  arXiv:1710.10196}, 2017.

\end{thebibliography}
\bibliographystyle{IEEETran}
\end{document}


\title{Learning stochastic object models from medical~imaging measurements using Progressively-Growing~AmbientGANs \\
 (Supplementary Material)}
\author{Weimin Zhou,
        Sayantan Bhadra,
        Frank J. Brooks, 
        Hua Li,~%
        and Mark A. Anastasio
}
\maketitle

  \vspace{-0.5cm}
  \begin{figure}[H]
	\centering
	\includegraphics[width=0.9\linewidth]{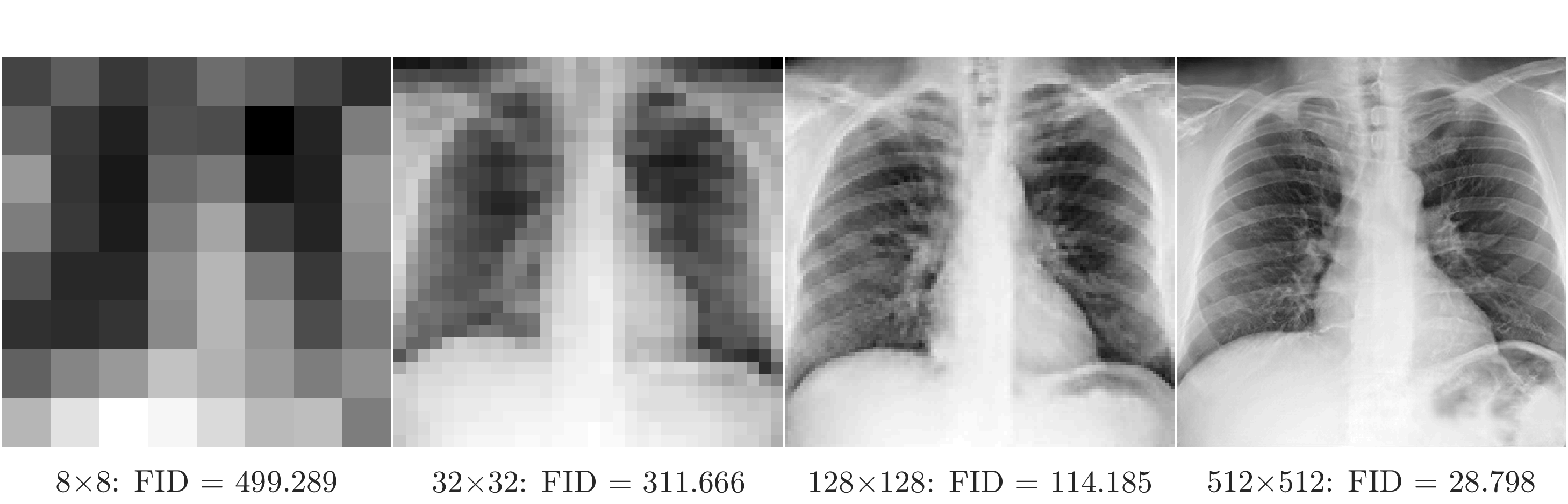}
	\caption{ProAmGAN-generated chest X-ray images at different training steps corresponding to Sec. \rom{4}-A in the manuscript. FID scores decreased as the resolution increased in the training process.}
	\label{fig:g_mri}
\end{figure}
  \vspace{-0.8cm}
  \begin{figure}[H]
	\centering
	\includegraphics[width=0.9\linewidth]{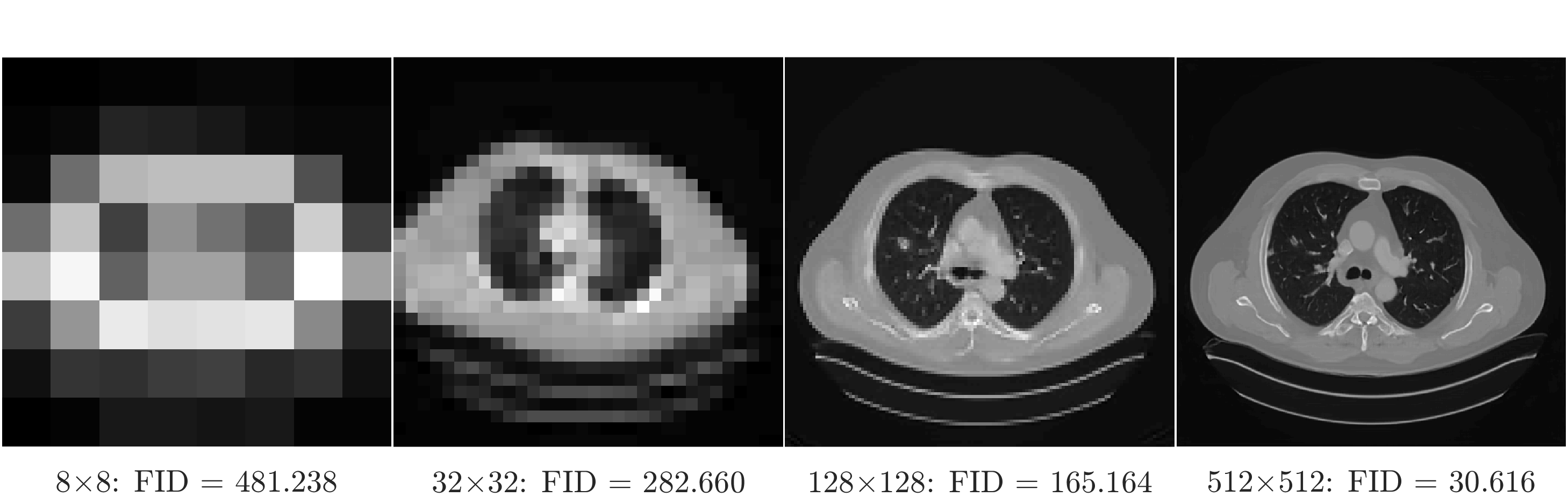}
	\caption{ProAmGAN-generated chest CT images at different training steps corresponding to Sec. \rom{4}-B in the manuscript. FID scores decreased as the resolution increased in the training process.}
	\label{fig:g_mri}
\end{figure}
  \vspace{-0.8cm}
  \begin{figure}[H]
	\centering
	\includegraphics[width=0.9\linewidth]{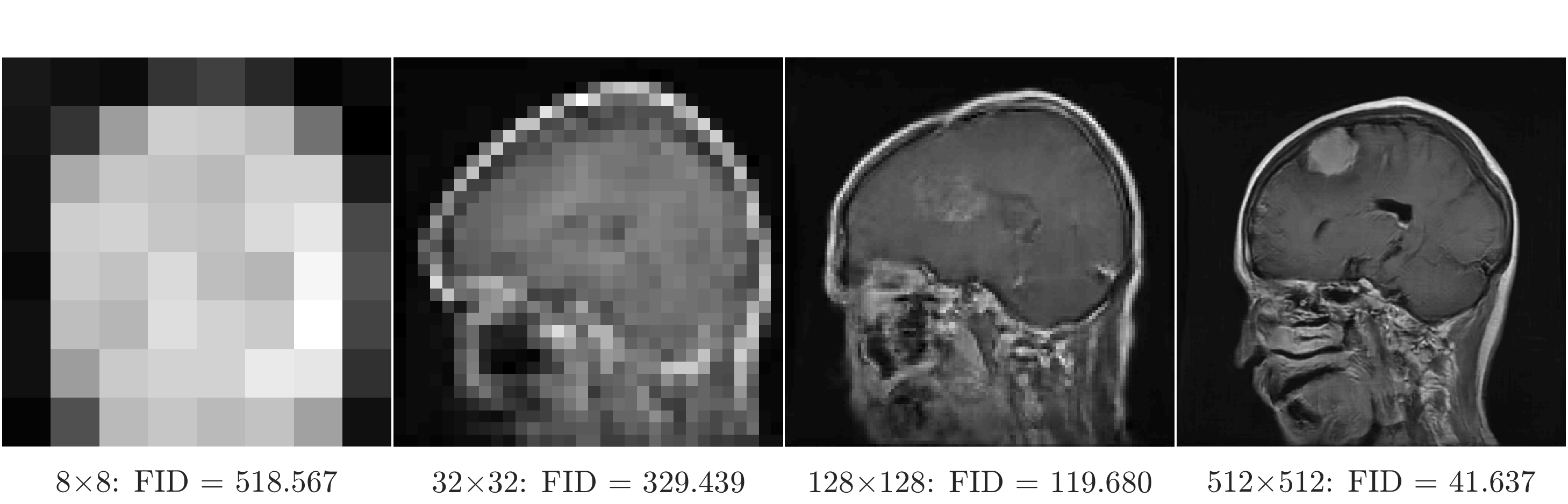}
	\caption{ProAmGAN-generated brain MR images at different training steps corresponding to Sec. \rom{4}-C in the manuscript. FID scores decreased as the resolution increased in the training process.}
	\label{fig:g_mri}
\end{figure}

  \vspace{-1cm}
  \begin{figure}[H]
	\centering
	\includegraphics[width=0.82\linewidth]{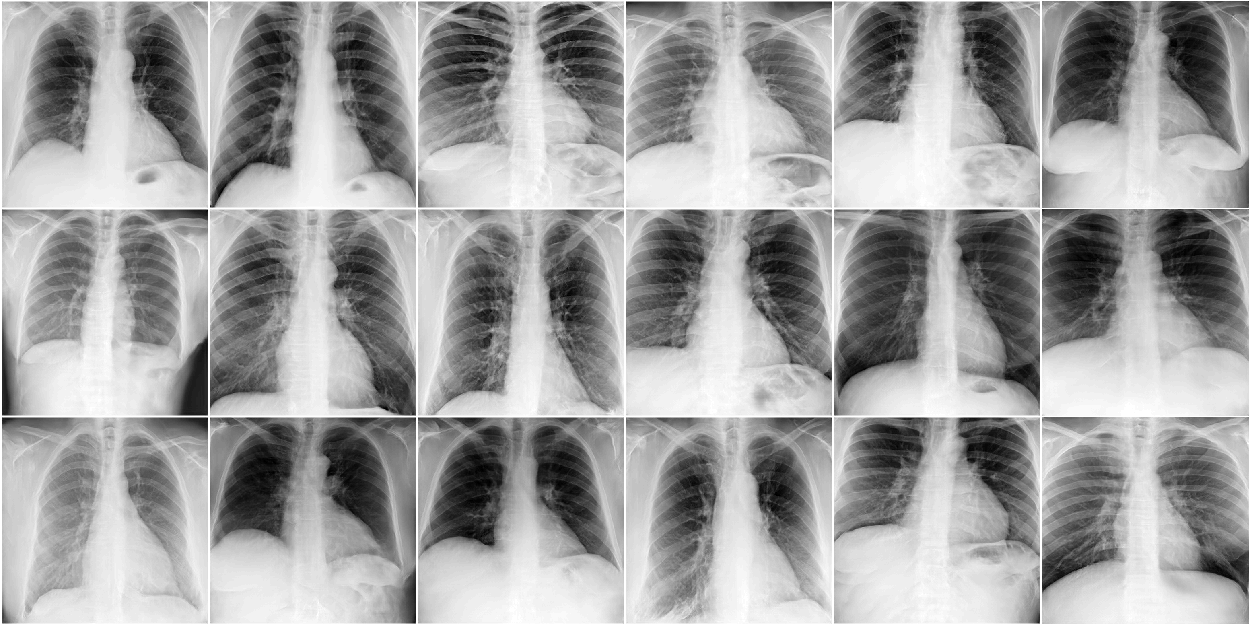}
	 \vspace{-0.1cm}
	\caption{Additional 18 ProAmGAN-generated chest X-ray images corresponding to Sec. \rom{4}-A in the manuscript.}
	\label{fig:g_mri}
\end{figure}
  \vspace{-0.5cm}
  \begin{figure}[H]
	\centering
	\includegraphics[width=0.82\linewidth]{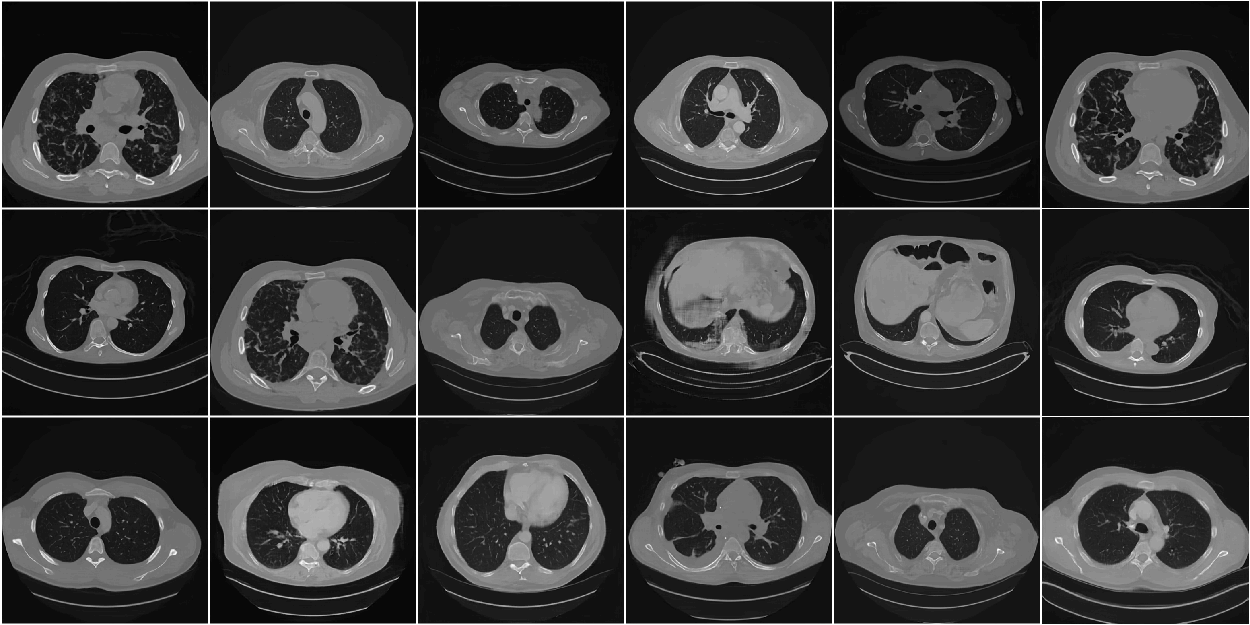}
	 \vspace{-0.1cm}
	\caption{Additional 18 ProAmGAN-generated chest CT images corresponding to Sec. \rom{4}-B in the manuscript.}
	\label{fig:g_mri}
\end{figure}
  \vspace{-0.5cm}
  \begin{figure}[H]
	\centering
	\includegraphics[width=0.82\linewidth]{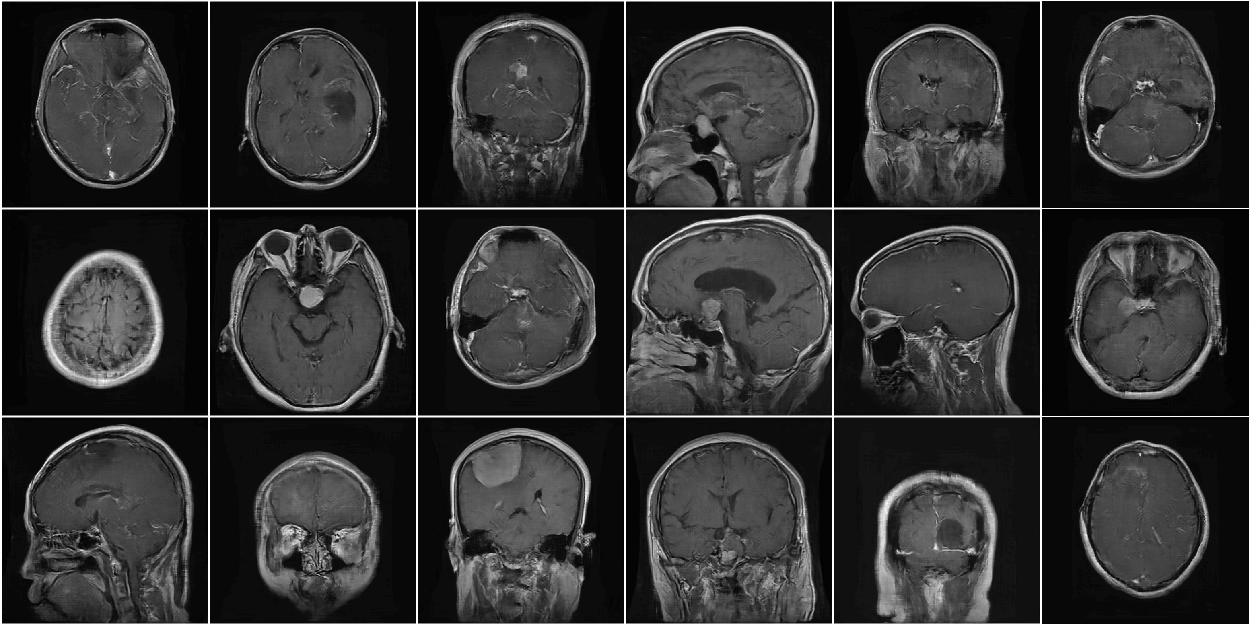}
	  \vspace{-0.1cm}
	\caption{Additional 18 ProAmGAN-generated brain MR images corresponding to Sec. \rom{4}-C in the manuscript.}
	\label{fig:g_mri}
\end{figure}

\vspace{-1.2cm}
\begin{figure}[H]
     \centering
 \includegraphics[width=0.6\linewidth]{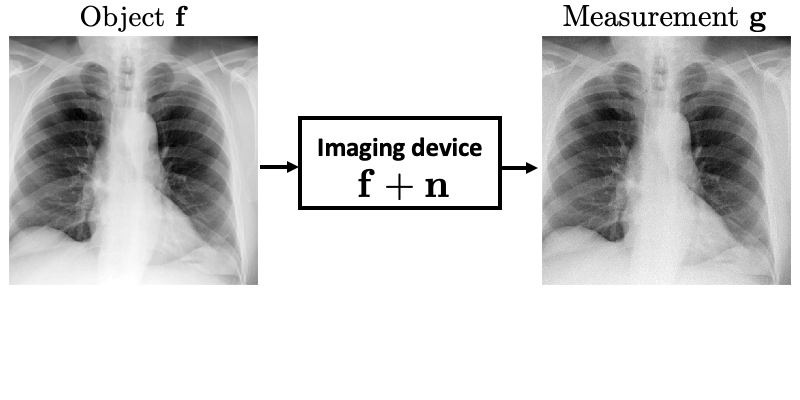}
 \vspace{-1.6cm}
 \caption{An illustration of idealized planar X-ray imaging system that acquires noisy imaging measurements.}
 \label{fig:g_x}
  \end{figure}

\vspace{-0.6cm}
\begin{figure}[H]
     \centering
 \includegraphics[width=0.6\linewidth]{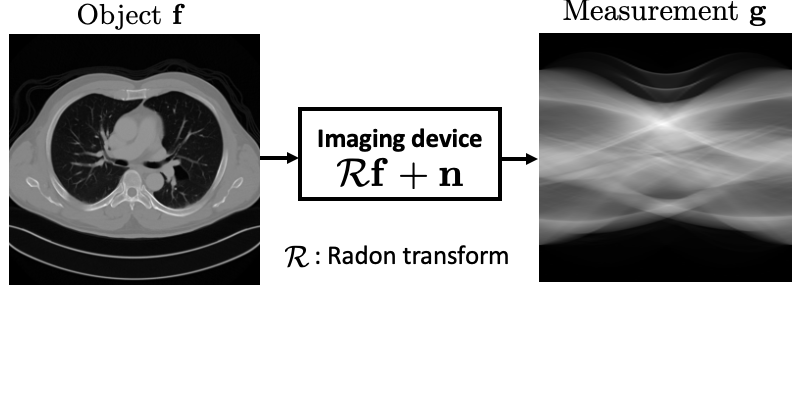}
 \vspace{-1.6cm}
 \caption{An illustration of tomographic imaging system that acquires Radon transform data.}
 \label{fig:g_ct}
  \end{figure}
  
  \vspace{-0.6cm}
  \begin{figure}[H]
	\centering
	\includegraphics[width=0.6\linewidth]{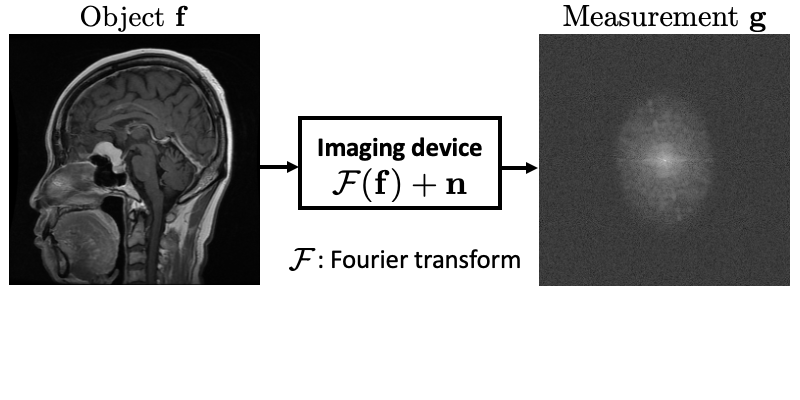}
	 \vspace{-1.6cm}
	\caption{MR imaging system with complete k-space data. Logarithm of one plus the magnitude of k-space data was displayed.}
	\label{fig:g_mri}
\end{figure}

\begin{table}[H]
\begin{minipage}[t]{0.48\linewidth}
\centering
\begin{adjustbox}{width=0.72\columnwidth,left}
\noindent\parbox[t][][t]{.74\linewidth}{
\begin{tabular}[t]{|lll|}
\hline
\textbf{Generator}       & Act.  & Output shape \\ \hline
Latent vector  & -      & 512$\times$1$\times$1      \\
Conv 4$\times$4      & LReLU & 512$\times$4$\times$4     \\
Conv 3$\times$3      & LReLU & 512$\times$4$\times$4      \\ \hline

Upscale          & -      & 512$\times$8$\times$8  \\
Conv 3$\times$3      & LReLU & 512$\times$8$\times$8      \\
Conv 3$\times$3      & LReLU & 512$\times$8$\times$8      \\ \hline

Upscale          & -      & 512$\times$16$\times$16  \\
Conv 3$\times$3      & LReLU & 512$\times$16$\times$16      \\
Conv 3$\times$3      & LReLU & 512$\times$16$\times$16      \\ \hline

Upscale          & -      & 512$\times$32$\times$32  \\
Conv 3$\times$3      & LReLU & 512$\times$32$\times$32      \\
Conv 3$\times$3      & LReLU & 512$\times$32$\times$32      \\ \hline

Upscale          & -      & 512$\times$64$\times$64  \\
Conv 3$\times$3      & LReLU & 256$\times$64$\times$64      \\
Conv 3$\times$3      & LReLU & 256$\times$64$\times$64      \\ \hline

Upscale          & -      & 256$\times$128$\times$128  \\
Conv 3$\times$3      & LReLU & 128$\times$128$\times$128      \\
Conv 3$\times$3      & LReLU & 128$\times$128$\times$128      \\ \hline

Upscale          & -      & 128$\times$256$\times$256  \\
Conv 3$\times$3      & LReLU & 64$\times$256$\times$256      \\
Conv 3$\times$3      & LReLU & 64$\times$256$\times$256      \\ \hline

Upscale          & -      & 64$\times$512$\times$512  \\
Conv 3$\times$3      & LReLU & 32$\times$512$\times$512      \\
Conv 3$\times$3      & LReLU & 32$\times$512$\times$512      \\ 
Conv 1$\times$1      & linear & 1$\times$512$\times$512      \\ \hline
\end{tabular}
}\hfill
\parbox[t][][t]{.4\linewidth}{
\begin{tabular}[t]{|lll|}
\hline
\textbf{Discriminator}       & Act.  & Output shape \\ \hline
Input image  & -      & $1\times512\times512$      \\
Conv $1\times1$      & LReLU & $32\times512\times512$      \\
Conv $3\times3$      & LReLU & $32\times512\times512$      \\ 
Conv $3\times3$      & LReLU & $64\times512\times512$     \\
Downscale & -      & $64\times256\times256$  \\ \hline

Conv $3\times3$      & LReLU & $64\times256\times256$      \\
Conv $3\times3$      & LReLU & $128\times256\times256$      \\ 
Downscale & -      & $128\times128\times128$  \\ \hline

Conv $3\times3$      & LReLU & $128\times128\times128$      \\
Conv $3\times3$      & LReLU & $256\times128\times128$      \\ 
Downscale & -      & $256\times64\times64$  \\ \hline

Conv $3\times3$      & LReLU & $256\times64\times64$      \\
Conv $3\times3$      & LReLU & $512\times64\times64$      \\ 
Downscale & -      & $512\times32\times32$  \\ \hline

Conv $3\times3$      & LReLU & $512\times32\times32$      \\
Conv $3\times3$      & LReLU & $512\times32\times32$      \\ 
Downscale & -      & $512\times16\times16$  \\ \hline

Conv $3\times3$      & LReLU & $512\times16\times16$      \\
Conv $3\times3$      & LReLU & $512\times16\times16$      \\ 
Downscale & -      & $512\times8\times8$  \\ \hline

Conv $3\times3$      & LReLU & $512\times8\times8$      \\
Conv $3\times3$      & LReLU & $512\times8\times8$      \\ 
Downscale & -      & $512\times4\times4$  \\ \hline

Minibatch stddev &- & $513\times4\times4$  \\ 
Conv $3\times3$      & LReLU & $512\times4\times4$      \\
Conv $4\times4$      & LReLU & $512\times1\times1$      \\ 
Fully-connected & linear      & $1\times1\times1$  \\ \hline
\end{tabular}
\\
\\
(a)
}
\end{adjustbox}
\end{minipage}
\hspace{-0.95cm}
\begin{minipage}[t]{0.48\linewidth}
\centering
\begin{adjustbox}{width=0.72\columnwidth,center}
\noindent\parbox[t][][t]{0.74\linewidth}{
\begin{tabular}[t]{|lll|}
\hline
\textbf{Generator}       & Act.  & Output shape \\ \hline
Latent vector  & -      & 512$\times$1$\times$1      \\
Conv 4$\times$4      & LReLU & 512$\times$4$\times$4     \\
Conv 3$\times$3      & LReLU & 512$\times$4$\times$4      \\ \hline

Upscale          & -      & 512$\times$8$\times$8  \\
Conv 3$\times$3      & LReLU & 512$\times$8$\times$8      \\
Conv 3$\times$3      & LReLU & 512$\times$8$\times$8      \\ \hline

Upscale          & -      & 512$\times$16$\times$16  \\
Conv 3$\times$3      & LReLU & 512$\times$16$\times$16      \\
Conv 3$\times$3      & LReLU & 512$\times$16$\times$16      \\ \hline

Upscale          & -      & 512$\times$32$\times$32  \\
Conv 3$\times$3      & LReLU & 512$\times$32$\times$32      \\
Conv 3$\times$3      & LReLU & 512$\times$32$\times$32      \\ \hline

Upscale          & -      & 512$\times$64$\times$64  \\
Conv 3$\times$3      & LReLU & 256$\times$64$\times$64      \\
Conv 3$\times$3      & LReLU & 256$\times$64$\times$64      \\ \hline

Upscale          & -      & 256$\times$128$\times$128  \\
Conv 3$\times$3      & LReLU & 128$\times$128$\times$128      \\
Conv 3$\times$3      & LReLU & 128$\times$128$\times$128      \\ \hline

Upscale          & -      & 128$\times$256$\times$256  \\
Conv 3$\times$3      & LReLU & 64$\times$256$\times$256      \\
Conv 3$\times$3      & LReLU & 64$\times$256$\times$256      \\
Conv 1$\times$1      & linear & 1$\times$256$\times$256      \\ \hline
\end{tabular}
}\hfill
\parbox[t][][t]{.4\linewidth}{
\begin{tabular}[t]{|lll|}
\hline
\textbf{Discriminator}       & Act.  & Output shape \\ \hline
Input image  & -      & $1\times256\times256$      \\
Conv $1\times1$      & LReLU & $64\times256\times256$      \\
Conv $3\times3$      & LReLU & $64\times256\times256$      \\
Conv $3\times3$      & LReLU & $128\times256\times256$      \\ 
Downscale & -      & $128\times128\times128$  \\ \hline

Conv $3\times3$      & LReLU & $128\times128\times128$      \\
Conv $3\times3$      & LReLU & $256\times128\times128$      \\ 
Downscale & -      & $256\times64\times64$  \\ \hline

Conv $3\times3$      & LReLU & $256\times64\times64$      \\
Conv $3\times3$      & LReLU & $512\times64\times64$      \\ 
Downscale & -      & $512\times32\times32$  \\ \hline

Conv $3\times3$      & LReLU & $512\times32\times32$      \\
Conv $3\times3$      & LReLU & $512\times32\times32$      \\ 
Downscale & -      & $512\times16\times16$  \\ \hline

Conv $3\times3$      & LReLU & $512\times16\times16$      \\
Conv $3\times3$      & LReLU & $512\times16\times16$      \\ 
Downscale & -      & $512\times8\times8$  \\ \hline

Conv $3\times3$      & LReLU & $512\times8\times8$      \\
Conv $3\times3$      & LReLU & $512\times8\times8$      \\ 
Downscale & -      & $512\times4\times4$  \\ \hline

Minibatch stddev &- & $513\times4\times4$  \\ 
Conv $3\times3$      & LReLU & $512\times4\times4$      \\
Conv $4\times4$      & LReLU & $512\times1\times1$      \\ 
Fully-connected & linear      & $1\times1\times1$  \\ \hline
\end{tabular}
\\
\\
\\
\\
\\
(b)
}
\end{adjustbox}
\end{minipage}
\caption{The architectures of the generator and discriminator for generating $512\times 512$ images (a) and those for generating $256\times 256$ images (b). 
More details about each component in the architecture can be found in ProGAN paper\cite{karras2017progressive}.}
\end{table}
  \vspace{-0.5cm}
\bibliography{PAmbientGAN}{}
\bibliographystyle{IEEETran}